# How Does the Adoption of Ad Blockers Affect News Consumption?

Shunyao Yan, Klaus M. Miller, Bernd Skiera†

Goethe University Frankfurt

July, 2021

**Abstract:** Ad blockers allow users to browse websites without viewing ads. Online news providers that rely on advertising revenue tend to perceive users' adoption of ad blockers purely as a threat to revenue. Yet, this perception ignores the possibility that avoiding ads—which users presumably dislike—may affect users' online news consumption behavior in positive ways. Using 3.1 million anonymized visits from 79,856 registered users on a news website, we find that adopting an ad blocker has a robust positive effect on the quantity and variety of articles users consume (21.5% - 43.3% more articles and 13.4% - 29.1% more content categories). An increase in repeat user visits of the news website, rather than the number of page impressions per visit, drives the news consumption. These visits tend to start with direct navigation to the news website, indicating user loyalty. The increase in news consumption is more substantial for users who have less prior experience with the website. We discuss how news publishers could benefit from these findings, including exploring revenue models that consider users' desire to avoid ads.

---

† Department of Marketing, Faculty of Business and Economics, Theodor-W.-Adorno-Platz 4, 60323 Frankfurt am Main. Email: yan@wiwi.uni-frankfurt.de; klaus.miller@wiwi.uni-frankfurt.de; skiera@wiwi.uni-frankfurt.de.

Ad blockers are software programs, usually browser extensions, that internet users can download to block advertising. They are highly popular, with recent reports suggesting that almost 50% of internet users between the ages of 16–64 are likely to have used an ad blocker within a given month (Kemp 2020). Despite their popularity, little empirical knowledge exists about how ad blockers impact advertisers, publishers, i.e., websites offering ad slots, or users. What is clear, however, is that neither publishers nor advertisers like ad blockers. Representatives of the advertising industry have described ad blockers as plain “robbery” and an “existential threat to the industry” (IAB 2015), and publishers have (unsuccessfully) sued providers of ad blockers for anti-competitive conduct, copyright infringement, and all kinds of unethical business practice (Toulas 2019).

The industry pushback against ad blockers exposes a potentially problematic assumption underlying the advertising revenue model—an assumption rarely explicitly discussed in articles on such models (e.g., Lambrecht et al. (2014)). Specifically, the advertising model assumes that users will be willing to ‘pay’ the publisher for the content they consume by enduring exposure to ads. However, the popularity of ad blockers suggests that, for many users, the price is too high: Users do not like ads, and indeed, this dislike seems to be the primary motivation for using an ad blocker (Sołtysik-Piorunkiewicz et al. 2019; Vratonjic et al. 2013).

Herein, we provide publishers with a more nuanced and potentially more positive view of ad blockers by taking an in-depth look at how adopting an ad blocker affects users’ behavior on a publisher’s website. In particular, we suggest that, in light of users’ dislike of ads, their adoption of an ad blocker might lead them to behave differently than they would do when exposed to ads. These behaviors could be positive from the publisher’s perspective. For example, exposure to annoying and distracting ads might interfere with users’ news consumption on a news site. In such a case, ad blockers could encourage users to consume and engage with more of the publisher's content.

To explore these ideas, we exploit a unique individual-level panel dataset from a reputed news website containing the data of 3.1 million visits of 79,856 registered users. Our dataset provides information about users’ news reading behavior, their usage of ad blockers, and the timing of their adoption (or dis-adoption) of ad blockers. We use this dataset to empirically examine the effects of ad blocker adoption on users’ news consumption, measured by the quantity and the variety of news consumption. Our analysis incorporates numerous robustness checks, including multiple alternative definitions of treatment. We further explore potential mechanisms for these effects by investigating users’ behavior within and across visits. We evaluate the heterogeneity of the effects across different news categories and different subsets of users (e.g., users with varying levels of user experience with our news website).

Overall, our findings contribute towards a better understanding of the effects of ad blockers on users’ behavior and reveal some of the benefits of ad blockers to publishers. Accordingly, they assist publishers in identifying alternative revenue models beyond the advertising model that could be more suitable for the news consumption context.

## *LITERATURE REVIEW*

Our study draws from and contributes to three main streams of literature.

First, we contribute to the understanding of the *effects of ad blocking.* A recent review paper identified ad blockers as one of the four primary sources of market inefficiency in digital advertising that warrant further research (Gordon et al. 2020). Preceding internet ad blocking, prior work analyzed consumer ad avoidance behavior in different media consumption contexts. Empirical evidence on ad avoidance focused on TV consumption and documented the existence of ad skipping (Danaher 1995) and its impact on consumer shopping behavior (Bronnenberg et al. 2010), and how to optimize the pricing and scheduling of TV advertising to correct for the loss in the audience due to ad skipping (Wilbur et al. 2013).

Among the studies that address internet ad blocking, most take an analytical approach (Aseri et al. 2020; Despotakis et al. 2021), whereas only a few rely on empirical evidence. These empirical studies include the work of Shiller et al. (2018), who used aggregate website-level data to show that an increase in ad blocker users reduces the website's traffic through a decrease in investment in content over three years. Another study by Todri (2021) used consumer-level data to show that ad blocking decreases online purchasing and reduces consumer search across information channels.

Our study contributes to the literature on ad blockers by using an elaborate real-world dataset to empirically investigate the impact of individual-level ad blocker adoption on news consumption—a context in which an understanding of the effects of ad blockers can be of practical value, given the online news industry's transition from advertising- to subscription-revenue models.

In doing so, our research contributes to a second, growing stream of literature, encompassing an ongoing debate about the *challenges faced by the news industry in the digital world*. This stream of literature has analyzed how social media (Scharkow et al. 2020; Schmidt et al. 2017), news aggregators (Calzada and Gil 2020; Chiou and Tucker 2017), other referral channels (Bar-Gill et al. 2021; Roos et al. 2020), and digital subscription models (e.g., paywalls and subscription fees; Lambrecht and Misra (2017); Aral and Dhillon (2021)) impact the demand for news. We contribute to this literature by analyzing the impact of ad blockers, which have not been studied in the context of news consumption. Notably, studies that examine advertising in conjunction with news consumption have tended to assume that exposure to advertising does not affect users' engagement with the news content itself (Aribarg and Schwartz 2019; Pattabhiramaiah et al. 2018). Our study challenges this assumption, enabling us to question whether the vast popularity of advertising-revenue models in the news industry is justified.

In comparing adoption versus non-adoption of ad blockers, we effectively compare different levels of exposure to advertising. This comparison enables us to contribute to the third stream of literature, the broad literature on *the effects of advertising*. Although advertising has been studied extensively in marketing and economics, prior studies have tended to focus on the success of advertising, as captured by measures such as recall and recognition of ads (Aribarg and Schwartz 2019), click-through (Dinner et al. 2013), view-through (Chae et al. 2019), sales (Danaher and Dagger 2013), and brand awareness (Bruce et al. 2020). These papers analyze the effectiveness of advertising on different media (Danaher and Dagger 2013) and, more recently, the potential negative marginal return of repetitive advertising exposure (Chae et al. 2019).

Much less research has examined how ads affect the platforms that publish them and users' engagement with those platforms. Among the empirical studies that examine these facets, most focus on non-digital markets (e.g., traditional TV, magazines, and yellow page books), documenting both positive and negative effects of ads on media consumption (Kaiser and Song 2009; Rysman 2004; Wilbur et al. 2013). Studies on digital advertising, in turn, have primarily taken place in highly controlled lab settings and thus are of limited relevance to real-world consumption. These studies have primarily revealed adverse effects of exposure to advertising on website usage (see, e.g., Goldstein et al. (2014)), but real-world evidence is lacking.

## *EMPIRICAL SETTING, DATASET, AND VARIABLE CONSTRUCTION*

We rely on a proprietary dataset from a reputable European news publisher who prefers to remain anonymous. In addition to a printed newspaper, our news publisher runs an online news website that publishes daily news, focusing on politics and business while reporting various other topics. The news website ranks among the top 10 in its country in weekly usage. At the time of our study, around 78% of the traffic to the news website came from its own country. Our news publisher has long been regarded as a national "newspaper of record" in the industry. Its reputation in its linguistic area is comparable to that of the *New York Times,* the *Financial Times,* or the *Guardian*. During our observation period, the news website offered all content free of charge. Users were required to register with the website (i.e., to enter their email addresses) to access archival content and newsletters but were not required to pay for this content. Approximately 20% of visits to the website came from registered users.

Our dataset was composed of clickstream data for all registered (but anonymized) users who visited the news website from the second week of June 2015 (week 1) to the last week of September 2015 (week 16). We focus on registered users because we can track them at the individual-level over time, which provides us with a unique panel setting. The clickstream data for each registered user included a complete record of that user's browsing behavior (including, among others, the time stamp, the page views, and whether an ad blocker is used) on the news website throughout the data collection period. We further combined the clickstream data with self-reported user demographics from the publisher's CRM database. In total, we analyzed 79,856 unique users with 3.1 million visits.

*Information about the Ads on the News Website*

Before analyzing how ad blocker adoption affected news consumption on our website, it is essential to acknowledge how ad blockers might have affected users' browsing experience. In most cases, an ad blocker automatically removes all ads on the webpage the user visits—except for websites that the user has whitelisted. Some websites require users to disable their ad blockers to view content; neither our website nor its main competitors did so during the observation period. Since the internet extensively displays ads, installing an ad blocker can lead to a noticeable difference in the display of a website, as shown in Figure 1, and the respective user experience.

[Insert Figure 1 about here]

Our news website runs display advertising according to the standard advertising formats outlined by the Interactive Advertising Bureau (IAB 2017). More precisely, our website runs leaderboard ads (728 × 90 pixels) on top of the page and rectangle ads (300 × 250 to 336 × 280 pixels) in the middle of the page on both desktop and mobile devices. Also, our website runs skyscraper ads (120 × 600 pixels) on the side of the page on desktops. On average, there are five display ads on the homepage and three display ads on an article page. These levels of advertising are comparable to, and in some cases even lower than, the levels of advertising on other similar premium news websites such as the *New York Times*, the *Washington Post,* and the *Guardian*. We further note that our news website does not run large half-page ads (300 × 600 pixels) or large mobile banner ads (320 × 100 pixels), whose removal by an ad blocker could lead to substantial changes in the display of the content. In addition, our website did not run native advertising during the observation period of our study.

*Ad Blocker Adoption Decisions and Trends in Ad Blocker Usage*

Crucial to our identification strategy is that an ad blocker is adopted on the browser and, thus, does not target to remove ads for a specific website. Surveys also suggest that a user's decision to adopt an ad blocker is motivated by ad annoyance, page loading speed, and privacy concerns (Sołtysik-Piorunkiewicz et al. 2019; Vratonjic et al. 2013). In addition, users may learn accidentally about the existence of ad blockers at different points in time.

Our observation period included an exogenous event that likely increased consumer awareness and, thus, ad blocker adoption: On September 16, 2015, Apple updated its mobile operating system and incorporated the ad blocking feature for the first time. This update was followed by a substantial increase in users' interest in ad blockers, as reflected in a sharp spike in the number of Google searches for the term "ad blocker" in September 2015, as shown in Figure S1 in the Web Appendix. We leverage this external event to run a robustness analysis based on a Heckman selection model (see Web Appendix G); the results of this analysis suggest that nothing specific about a user's news consumption influences their decision to adopt an ad blocker.

*Construction of Independent Variables – Ad Blocker Adoption*

Our dataset records, for each user, the number of page impressions with blocked ads. We use this information to derive an indicator of ad blocker usage. Specifically, a positive number of page impressions with blocked ads indicates an ad blocker's usage, whereas zero impressions blocked indicates no usage. Of our 79,856 users, 19,088 users used an ad blocker during this period (as indicated by a positive number of ad impressions blocked), and 60,768 users did not. Thus, 24% of users in our dataset used an ad blocker; this percentage is comparable to the ad blocker adoption rates across European countries during that period, ranging from 20% to 38% (Newman et al. 2016).

Whereas our dataset contains records of news consumption from week 1, it only recorded ad blocker usage from week 10 onwards, which creates a potential left-censoring problem. For example, a user with ad blocker usage in week 10 might have used it in week 9 or earlier. Thus, such a user would not experience a change in her exposure to ads. Following an approach to address a similar problem in the context of adopting Spotify (Datta et al. 2018), we designate a two-week cut-off period (e.g., weeks 10 and 11[1]) for defining our group of ad blocker adopters. Precisely, we classify users who have zero ad blocker usage in weeks 10 and 11 (i.e., the first two weeks of observed ad blocker usage) and then have non-zero ad blocker usage from week 12 as ad blocker adopters in week 12. Thus, these ad blocker adopters constitute the primary treatment group for our analysis. In addition, we classify a user into the control group, in turn, if the user had no ad blocker usage throughout weeks 10–16. We do not include users who did not visit our website during cut-off weeks (e.g., weeks 10–11) in the analysis[2].

According to this construction, our treatment group comprises 6,366 users. Our control group comprises 38,270 users (see the top part of Figure 2 to illustrate our construction of the treatment and control groups).

For robustness, we carry out two additional analyses using alternative definitions of treatment and control groups. Our second analysis compares early adopters of an ad blocker (treatment group) to late adopters (control group). We define an early adopter as a user who adopts an ad blocker in week 12 (*n* = 1,124) and a late adopter as a user who adopts an ad blocker in week 14 (*n* = 1,167). This approach enables us to control for bias related to adopting an ad blocker because both treatment and control groups adopted an ad blocker and only differ in the timing at which they did so.

---

[1] We show the robustness of the cut-off threshold by reducing it to 1 week and extending it to 4 weeks in Web Appendix D.
[2] Users who visited our website during weeks 10-11 but did not return to the website were not considered here because they are pruned out later in the matching process (see Table 3: the last observed week starts from week 13).

Our third analysis leverages the fact that, in our sample, 9,055 of the total 79,856 users had already adopted an ad blocker during our cut-off periods (weeks 10 and 11). Thus, they are censored users for whom we cannot identify the timing of adopting an ad blocker. However, we can identify the timing at which they dis-adopted an ad blocker based on their non-zero to zero change in ad blocker usage for these users. Therefore, we use these censored adopters to identify the effect of the abandonment of ad blockers. Precisely, we classify these censored adopters into the treatment group if they change from non-zero to zero ad blocker usage during week 12 or later. In other words, the users in this treatment group (*n* = 2,882) do not see ads in weeks 10 and 11 but start to see ads during week 12 or later. The control group (*n* = 6,173) consists of users in this censored sample who have non-zero ad blocker usage throughout weeks 10–16. The bottom part of Figure 2 depicts the construction of these treatment and control groups.

[Insert Figure 2 about here]

*Construction of Dependent Variables*

Our analysis considers numerous news consumption and users' behavior measures, which we summarize in Table 1. We report all measures and the corresponding analyses at the user-week level.

Our primary variables of interest are (i) article views, i.e., a count of the number of news articles a user clicks on the news website; this measure captures the *quantity* of news consumption; and (ii) breadth, a count of the number of unique news categories of article views; this measure captures the *variety* of news consumption.

We further explore mechanisms that might underlie the effects of ad exposure on our main variables of interest. For this analysis, we measure variables corresponding to (i) users' behavior within visits—including, for example, the number of article views per visit and time spent on the website per visit; and (ii) users' behavior across visits, including, e.g., the total number of visits, as well as information about the referral sources of these visits (e.g., direct navigation to the news website versus referral from external websites such as social media platforms or search engines). A visit is an entry to our news website that ends with the user clicking away from the website or remaining inactive for 30 minutes. We elaborate the complete list of within- and cross-visit variables in Table 1.

We base our mechanism analysis on the premise that adopting an ad blocker is likely to affect users' behavior through two main channels. The first is a cognitive mechanism, wherein the absence (as opposed to presence) of ads resulting from ad blocker adoption enhances the user's cognitive resources to process website content. We suggest that a cognitive mechanism is likely to manifest in short-term, within-visit effects, such as an adoption-induced increase in the number of article views per visit. The second potential mechanism is a learning mechanism. User experience with an ad blocker helps users learn how well their preferences match the website and increase their desire to return to the website. This mechanism manifests in the repeat visits and especially direct navigation to the website instead of referring channels.

*Additional variables.* To obtain a more detailed understanding of the main effects observed and establish robustness, we also count the number of article views separately for each news category (e.g., political news and economic news). We report all news categories of our website in Panel 2 of Table 1. In addition, we count the number of page views on the home page. The page views from news articles and the home page account for more than 90% of the browsing behavior during our observation period. Other browsing behaviors on our news website include browsing account-related pages and the weather forecast.

[Insert Table 1 about here]

*Descriptive Statistics and Preliminary Analysis*

Recall that our primary analysis is on two groups of users: ad blocker adopters (treatment group) and non-ad blocker adopters (control group). We report in Table 2 the group means of our main dependent variables: article views and breadth, a before-and-after difference within each of the two groups, and a difference-in-differences (DiD) comparison between groups.

We start with the before-and-after analysis. For treated users (ad blocker adopters), all news consumption measures increase after treatment. However, these measures decrease for the control group (non-ad blocker adopters), leading to a positive value when we compute a simple DiD estimator (e.g., +1.89 for article views and +0.62 for breadth; see the last column of Table 2 for all adopters). These results provide preliminary evidence of the positive effect of ad blocker adoption on news consumption in terms of both quantity and variety. However, self-selection into ad blocker adoption could confound the DiD reported in Table 2.

[Insert Table 2 about here]

In the next section, we describe our approach to remove any potential confounders from self-selection. To identify the causal effect of ad blocker adoption on news consumption, we combine matching with DiD, and establish robustness by repeating the analysis with the alternative treatment and control definitions outlined above.

## *IDENTIFICATION STRATEGY*

Selection bias into treatment can come from both observable and unobservable confounders. In our analysis, we first non-parametrically control for observable confounders using coarsened exact matching (CEM). Then, to remove any time-invariant unobserved confounders, we use DiD with individual-level fixed effects. As for time-varying confounders, we use a placebo treatment test to show that they do not bias our results. In what follows, we describe the identification strategy for our primary analysis, in which ad blocker adoption at week 12 and onward (as outlined above) serves as a treatment. We use a similar strategy in our two robustness analyses, in which, respectively, early adoption and dis-adoption of an ad blocker serve as treatment (see above).

Recall that our sample covers 16 weeks, running from June 8, 2015, to September 27, 2015. Our treatment starts from week 12. We use the first 11 weeks, i.e., the entire pre-treatment period, for matching. For our estimation, we use weeks 7 to 11 as the pre-treatment period and the remaining weeks, weeks 12 to 16, as the post-treatment period. We also use weeks 1 to 11 as a pre-treatment period in an additional robustness check, reported in Web Appendix Table S10, which confirms the robustness of our results.

*Coarsened Exact Matching*

To remove observable confounders, we use matching to refine our treatment and control groups. The statistical treatment effects literature commonly combines matching methods with DiD (Heckman et al. 1998). We chose CEM for its advantages over other matching methods (such as propensity score matching; see King and Nielsen (2019)) in terms of balancing the covariates (Iacus et al. 2012) when the number of covariates is not large. We also use propensity score matching as a robustness check (Web Appendix Table S6). The results remain similar.

CEM is a nonparametric method of controlling for observed confounders (more commonly called "covariates" in the matching literature; herein, the two terms are used interchangeably). CEM involves matching on a vector of covariates (instead of on a scalar representing a distance metric summarizing all covariates, as in other matching methods) and using the covariates' original dimensions. Categorical covariates can be kept at their original values. In contrast, continuous covariates are 'coarsened' into bins (e.g., the number of page views per week, which may range between 1 and 50, can be coarsened into the bins 1 to 5, 6 to 10, etc.). Then, CEM conducts exact matching with all covariates, with the continuous covariates coarsened.

Each observation belongs to a unique stratum containing all observations with identical values of all covariates (e.g., a male who is 30-35 years old, who views 11-15 pages a week, is placed in a stratum containing other males who are 30-35 years' old who view 11-15 pages a week). CEM keeps all treated and control units within a stratum (as all observations have the same values for the covariates within a stratum) and adjusts the unequal number of matched treated and control units with weights (which we further use in our regression analysis). Strata containing zero-treated units or zero-control units are pruned, and the remaining strata are weighted according to the number of observations they contain for further analysis.

In this way, CEM balances the covariates on their original dimensions and eliminates differences in covariates between the treatment and control group in all moments, quantiles, and functional forms (Iacus et al. 2012). In contrast, other matching methods that focus on the univariate balance of the means of covariates (e.g., through estimating and matching on a propensity score obtained from a logit regression on covariates) might not remove, and in fact, can even increase, bias due to imbalances of other moments or functional forms (King and Nielsen 2019).

We can then combine CEM with other causal inference methods, which normally require a model. In these cases, the application of CEM (as opposed to reliance on a full, non-matched sample or a sample matched by another model-dependent matching method) can help decrease model dependence and statistical bias (Iacus et al. 2012; Zervas et al. 2017). However, when the number of covariates gets large, it can be challenging to find exactly matched pairs, even with coarsened variables. In this case, matching based on a univariate score (such as propensity score) is more efficient. In addition, a univariate score also helps in terms of visualization, as we did in Figure 3.

We match the treatment and control groups based on the following covariates[3]. First, we include three controls for *demographics*—age, gender, and income—as prior empirical studies show that these demographics are important determinants of news consumption (Fan 2013; Roos et al. 2020) and ad blocker usage (Sołtysik-Piorunkiewicz et al. 2019; Zhao et al. 2017). Second, we match users on *pre-treatment browsing* behavior, namely, article views, breadth, and numbers of visits. Third, to make sure users are active throughout the same observation window and to reduce bias related to a user's visit behavior or active time, we include a user's *first and last observed week* in the matching process. Finally, we include the two variables that stay significant when we model the users' ad blocker adoption process (reported in Table S14 of Web Appendix G): *most frequently used browser* and *mobile page views in the pre-treatment period*.

[Insert Table 3 about here]

Table 3 compares the matched and unmatched samples of the treatment group (ad blocker adopters) and the control group (non-ad blocker adopters) regarding the observed characteristics that we use for matching. Before matching, we examine how user demographics impact the ad blocker adoption probability using logistic regression in Web Appendix K (Table S22). We find that the odds of being an ad blocker adopter in the male group is 1.33 times that of being an ad blocker in the female group and the odds of being an ad blocker adopter in the Income Index 2 group is 0.63 times that of being an ad blocker in the Income Index 1 group. The right part of Table 3 shows that, after matching, the treatment group and the control group are balanced in all the matching variables.

[3] We note the tradeoff in using demographic data for matching by introducing a selection problem, as those who report their demographics in CRM data can be different from those who do not report. However, in the Web Appendix Table S4, we confirm that demographic data are missing at random. We also report the estimation results using unmatched data as a robustness check. The results remain similar.

Figure 3 provides further validation to our matching process; specifically, it depicts the distributions of the 'propensity scores', the distance metric obtained by running a logistic regression with all matching variables, the treatment group, and the control group before and after CEM. Figure 3 shows that CEM balanced the treatment and control groups by creating a similar empirical distribution of the matching variables and thus increasing the common support (or overlap) between these two groups. Figure 3 also shows that CEM removes the most likely and the least likely ad blocker adopters from the sample. Thus, in effect, we mimic an experimental setting in which users in the treatment and control group are equally likely to adopt an ad blocker but decide at random whether to adopt it or not.

CEM removes the differences in all observed covariates, and thus any remaining selection bias can only come from unobservable confounders, which we discuss in the next section.

[Insert Figure 3 about here]

*Difference-in-Differences (DiD)*

Having produced our matched treatment and control samples, we subsequently apply DiD with individual-level fixed effects, removing time-invariant unobserved confounders by taking the temporal differences within each user. We eliminate all variation in news consumption caused by time-invariant unobserved heterogeneity between individuals (e.g., differences in education or preference towards news or ads). In addition, DiD removes any bias due to time trends that are common to both groups (e.g., resulting from seasonality or news shocks) by taking the difference once again across groups.

Specifically, we estimate the following DiD model:

(1) $$Y_{it} = \alpha_i + \delta_t + \beta_1 * I_{it1}(within\ 1\ week\ of\ Treatment_{it}) + \beta_2 * I_{it2}(remaining\ weeks\ since\ Treatment_{it}) + \varepsilon_{it},$$

where $Y_{it}$ is the dependent variable, one of the news consumption measures listed in Table 1, for user $i$ in week $t$; $\alpha_i$ is a user-fixed effect, controlling for time-constant differences across users, such as education or tastes towards news; $\delta_t$ is a week-fixed effect, controlling for common trends or changes over time that affect all users equally, such as breaking news; $I_{it1}$ is an indicator variable that is equal to one if the observation for individual $i$ in week $t$ is within one week after the treatment (so that this binary variable is one in the treated week and the following weeks); $I_{it2}$ is an indicator variable that is equal to one if the observation of individual $i$ in week $t$ is after one week post-treatment (so that this binary variable is 1 from week 2 to week 5 post-adoption); $\beta_1$ captures the effect for the treated week, i.e., the adoption week, and one week post-adoption; $\beta_2$ captures the effect for the remaining weeks (week 2 to week 5 post-adoption); and $\varepsilon_{it}$ is the error term, clustered at the user-level.

Our identification strategy builds upon the changes in news consumption after treatment (in our primary analysis, adopting an ad blocker). Crucial for this identification is that all confounders are either controlled for or quasi-random; that is, any unobserved time-varying confounders follow parallel trends in the pre-treatment periods. In the Web Appendix A, we formally test this identification condition and show that this identification assumption holds for all dependent variables using a placebo treatment test (Angrist and Pischke 2008).

## *RESULTS*

### *Main Effects on Quantity and Variety of News Consumption*

We are primarily interested in two measures of news consumption: the number of article views (i.e., the quantity of news consumption) and the number of news categories to which viewed articles correspond (i.e., the variety of news consumption). Panel 1 of Table 4 reports these results for our primary analysis and our two robustness analyses with alternative treatment and control group designs. The dependent variables are the natural logarithms of news consumption measures (with 1 added to correct for zero values); thus, a simple transformation of $\beta_i$ in the regression model (1) can be locally approximated as a percentage change in news consumption: $\exp(\boldsymbol{\beta_1}) - 1$ reports the effect during the week of treatment and the following week (referred to as a '1-week effect') and $\exp(\boldsymbol{\beta_2}) - 1$ reports a 5-week effect. However, as we add 1 to the dependent variable and our dependent variable Y is rather small, the percentage increase is constantly underestimated by $(1 - \frac{Y}{Y+1})$ %. We thus correct for this underestimation at the median value of the dependent variables (which we report in Table 1) by multiplying our percentage increase by $\frac{Y+1}{Y}$. We checked the robustness of the results regarding the decision to use a log-transformation and adding one to the dependent variable (see Web Appendix H) and found that the results remained consistent.

[Insert Table 4 about here]

Our primary analysis used ad blocker adopters as a treatment group and ad blocker non-adopters as a control group. We find a significant and consistent positive effect of ad blocker adoption on both the quantity and variety of news consumption. For the quantity of news consumption (see the first column in Panel 1 of Table 4), the number of article views increases by 21.5% (= (exp (0.155) – 1) * 1.25) at the median (4), lasting even 5 weeks after adoption. The increase is larger within the treated week and one week of the treatment: 43.3% (= (exp (0.297) – 1) * 1.25) at the median. The variety of news consumption, in turn, increases by 13.4% (= (exp (0.096) - 1) * 1.33) at the median (3) over 5 weeks, with a 1-week increase of 29.1% (= (exp (0.198) - 1) * 1.33) at the median.

Our second analysis, in which we compared early adopters (treatment group) who adopted an ad blocker at week 12 with late adopters (control group) who adopted at week 14, produced results consistent with those of our primary analysis. In this analysis, $\boldsymbol{\beta_1}$ measures the effect of ad blocker adoption on news consumption during the period in which early adopters had already adopted an ad blocker but late adopters had not (i.e., week 12 and week 13). $\boldsymbol{\beta_2}$ measures the effect of early adoption on news consumption over 5 weeks of post-treatment (i.e., week 12 to week 16). Notably, we do not expect to find a 5-week effect as in the previous analysis because the control group in this analysis adopts an ad blocker only two weeks after the treatment group's adoption.

In line with our expectations, we find a positive and significant effect during week 12 and week 13 and do not find any significant effect over 5 weeks. Specifically, early adopters increased the quantity of their news consumption by 49.2% (= (exp (0.332) - 1) * 1.25) at the median and increased the variety of their consumption by 30.6% (= (exp (0.207) - 1) * 1.33) at the median during week 12 and week 13. These estimates are numerically similar and stay within the standard error of those obtained in the previous analysis.

Our third and last complementary analysis, in which users who abandoned ad blockers served as the treatment group and users who used ad blockers throughout the entire observation period (weeks 10–16) served as a control group, further supports the robustness of our findings. Specifically, we find that users who abandoned ad blockers decreased their quantity of news consumption by 25.6% (= (exp (0.23) - 1) * 1.25) at the median and the variety of their news consumption by 19.2% (= (exp (0.135) - 1) * 1.33) at the median within 1 week of dis-adoption. This result only retains the sign that is consistent with the results of the previous two approaches, but not the significance level. One reason is that CEM matching process drops many observations. Indeed, in the Web Appendix E, we present a robustness check without matching (Table S12), in which we find that ad blocker abandoners significantly decrease their news consumption over 5 weeks.

Taken together, the findings of these three analyses support the positive effect of ad blocker adoption on both the quantity and the variety of news consumption.

*Effect of Ad Blocker Adoption on Quantity of News Consumption, by News Category*

Having established the effect of ad blocker adoption on the quantity and variety of news consumption, we decompose in Panel 2 of Table 4 the effect on article views into various news and non-news categories. These categories include ‘hard’ news (political news, economic news, and opinion news, following Angelucci and Cagé (2019)); ‘soft’ news (sports, culture & art, lifestyle news), and non-news article pages (e.g., account settings and play pages that include games such as Sudoku or Mahjong). For clarity of presentation, in what follows, we only report the results of our primary analysis, with ad blocker adopters as the treatment group and non-adopters as the control group. The results obtained with our alternative treatment definitions (early adoption of ad blockers or ad blocker abandonment) remain substantively the same, and we report them in the Web Appendix E.

We find that the increase in article views attributable to ad blocker adoption is driven primarily by increases in the consumption of hard news. None of the soft news categories has significant effects over 5 weeks, though there is a 1-week effect for Sports and Art & Culture. In addition, our analysis reveals no effect of ad blocker adoption on views of non-news pages. In the Web Appendix I, we report the complete results for each category (including science news, finance news, and news ticker, which Cagé (2020) also classified 'hard' news, i.e., news that affects the political process).

*Within-Visit Effects: Exploring a Cognitive Mechanism*

Previous studies have shown that ads have a cognitive impact on consumers, regardless of whether consumers pay attention to them (Vakratsas and Ambler 1999). The reason is that our brain processes information both consciously and subconsciously (Kahneman 1973). Thus, one potential explanation for the increases we observe in news consumption after adopting an ad blocker (e.g., as in Panel 1 of Table 4) may be the increased availability of cognitive resources attributable to not seeing as many ads. If this explanation holds, then we should expect to observe post-adoption changes in behavior within individual visits for two reasons. The cognitive system for processing information is the working memory that functions in the short-term (Baddeley 1992), and a visit is, by definition, a short-term period of engagement with the site (recall that a visit ends when the user actively clicks away or stays inactive for 30 minutes).

Panel 1 of Table 5 shows that the number of article views per visit increased by only 13.9% (= (exp (0.067) - 1) * 2) at the median within 1 week, which is much less than the 1-week increase of 43.3% of total article views (as reported in Panel 1 of Table 4). The number of other page views (e.g., homepage views) per visit also only increases by 7.7% (= (exp (0.047) - 1) * 1.6) at the median.

[Insert Table 5 about here]

Another means by which the availability of additional cognitive resources (attributable to ad blocker adoption) might influence users' news consumption behavior is by enabling them to read longer or more complex articles. To explore this possibility, we scraped the titles of all the articles (i.e., the headlines) that users clicked on and analyzed the length of these titles (by measuring the number of words). It serves users as an indicator for the expected length and complexity of the respective news article. We find the title length per visit increased by 18.6% (= (exp (0.146) - 1) * 1.18) at the median within 1 week while the title length per article did not increase significantly.

In addition, we checked the effect on time spent per visit, which is an engagement metric commonly known in the industry as dwell time (i.e., the time that a user spends on the website before clicking away). We find that the time spent per visit increased by 47.4% (= (exp (0.387) - 1) * 1) at the median over 1 week and by 24.3% (= (exp (0.217) - 1) * 1) at the median over 5 weeks after ad blocker adoption. This result suggests that ad blocker adoption might have enabled users to devote more attention to the articles they read, even though they did not read more news articles within each visit.

*Effects across Visits: Exploring a Learning Mechanism*

So far, these results provide the first suggestive evidence that ad blocker adoption elicited cognitive effects among users. Yet, these effects explain only a small part of the increase in the quantity of news consumption after the ad blocker adoption, suggesting that an increase in the number of visits to the website drives the major part of the increase. Indeed, we find both a 1-week increase and a 5-week increase in website visits (see column “Visits” in Panel 2 of Table 5). These effect sizes are comparable to those corresponding to the effect of the treatment on article views (Panel 1 of Table 4), which further supports the robustness of our main result.

Learning, the process of acquiring knowledge and experience about a product, provides an intuitive explanation for this increase in the number of visits (Johnson et al. 2003). In particular, the usage of ad blockers may have affected users’ experience of the site (e.g., by enhancing their enjoyment), thereby encouraging them to visit it more frequently.

Hoch and Deighton (1989) suggest that learning involves actively seeking experience with a product. To explore whether users engaged in an active information-seeking process, we separately analyzed visits to the news website of each referral source. Specifically, users could visit the news website directly (e.g., by using a bookmark or directly typing in the

URL) or could be referred by a social media website (e.g., Facebook), a search engine (primarily Google) or an email with a newsletter of the newspaper. We find that the users directly visiting the news website drive the increase in article views, indicating an active seeking process. It indicates that users enjoy the website experience when using an ad blocker and start to develop routines and habits associated with the website. Such routines and habits might last even longer than our observation period and represent an actual long-term effect, particularly given the inherently recurring nature of news consumption (DeFleur and Ball-Rokeach 1989).

Notably, our observation of an active information-seeking process can also explain the observed increase in the variety of news consumption: Users might have actively sought to experience more aspects of the news website, which may have led them to explore additional news categories.

Overall, these results enable us to conjecture regarding the mechanism of these effects, suggesting that they are more consistent with consumer learning (and less consistent with enhanced availability of cognitive resources due to reduction of ad exposure). However, the evidence on the underlying theoretical mechanism is only suggestive; conclusively distinguishing and identifying the mechanism is beyond the scope of this paper and would be an exciting avenue for future research.

*Heterogeneous Treatment Effects Across Users with Different Characteristics*

To obtain a more detailed understanding of the effects observed, we reran our analyses (regression (1)) while distinguishing among users according to specific characteristics of interest, namely the number of browsers used to access the website and the frequency of website usage before treatment. Table 6 presents the results of these analyses.

*Single- vs. multiple-browser usage.* In 2015, very few browsers blocked advertising by default. Instead, users had to install an ad blocker in their browsers and, when using multiple browsers (on a single device or across multiple devices), users had to install an ad blocker for each browser. In addition, not all browsers enabled an ad blocking feature (e.g., ad blocking was not feasible in the Apple mobile browser before September 2015, when iOS9 launched).

Our definition of ad blocker adoption included every user who has blocked at least one ad in the analyses described above. For users who use multiple browsers, however, this definition does not necessarily imply a total absence of exposure to ads; that is, a particular user might block ads on one browser yet view them on another. In such cases, the effect of ad blocker adoption should be weaker than in cases in which all ads are blocked. Our dataset enabled us to explore this idea, as it reveals which browser (including whether the device is a mobile or desktop) each user used to access our website; 54% of ad blocker adopters used multiple browsers. We reran our analyses, distinguishing between users who used multiple browsers and those who did not. As shown in Panel 1 of Table 6, we found that ad blocker adoption indeed had a larger effect on news consumption (both in terms of quantity and variety) for single-browser users than for multiple-browser users.

*Usage frequency.* We also shed light on how the effect differs across users with different prior experiences with the website. Previous studies found that heavy users are more likely to perform information-and variety-seeking behavior on the internet or online platform (Assael 2005; Gu et al. 2021). However, it is also more difficult for heavy users to increase their time on one platform even further. Our results show that light users (i.e., users whose frequency of visiting the website was below the median in the pre-treatment period) stronger increase their news consumption than heavy users (whose frequency of visiting the website was above the median) after adopting ad blockers. As shown in Panel 2 of Table 6, the effects were only significant within 1-week post-treatment for the heavy users. These results are also consistent with a potential learning mechanism: users who have less prior experience with the website are more strongly affected. Worth noticing, the light users in our analysis are registered users whose news consumption level is higher than that of non-registered users.

[Insert Table 6 about here]

## *SUMMARY AND CONCLUSION*

We used data from 3.1 million anonymized visits from 79,856 users on a news website to show that ad blocker adoption has a robust positive effect on news consumption. Further, we find that the increase in news consumption is driven by users visiting the news site more frequently (instead of reading more articles per visit), primarily through direct navigation to the website (instead of referral visits from social media).

This study offers several practical insights beyond contributing empirical evidence regarding the relationship between ad blocker adoption and news consumption. Our findings challenge the conventional assumption that users' adoption of ad blockers, which interferes with publishers' advertising revenue, is exclusively negative for publishers. Indeed, the increased engagement of ad blocker users would not directly translate into advertising revenue. However, ad blockers filter out more ad-sensitive users and leave the publishers with users who are more willing to get exposure to ads. Our analysis of the ad blocker adoption decision on user demographics shows that female users and high income users are less likely to adopt an ad blocker than male users and low income users. Thus, the users who remain exposed to advertising on our website have a more valuable demographic profile than the demographic profile of the users who are not exposed to advertising on our website (Lambrecht and Tucker 2019). Suppose publishers could further exploit this heterogeneity in ad sensitivity by selling subscriptions to ad blocker users and increasing the ad intensity for non-ad blocker users; they could achieve higher returns in revenue (Despotakis et al. 2021).

Furthermore, our results reveal that ad blocking increases users' consumption of news, in addition to enhancing other measures of engagement and retention with the publisher's website (e.g., frequency of visits). These enhanced engagements of ad blocker users could translate into subscription revenues, which more and more news publishers rely on. Our publisher introduced a paywall after our observation period. The paywall offered different subscription plans but still showed ads to all users independent of their subscription status. 1.5 years after the paywall introduction, the subscription rate of ad blocker users is 30.13% higher than the subscription rate of non-ad blocker users. This higher share outlines the potential of ad blocker users to contribute to the subscription revenue. Publishers also recognize that ad blocker users are usually more willing to pay for some kinds of subscriptions (Yeon 2020), which further supports our finding. In addition, we also found that ad blocker adoption had a stronger effect on the news consumption of light users. In this way, ad blockers can “convert” light users into heavy users, and heavy users are more likely to subscribe than light users (Anderson et al. 2020).

Another implication of our findings is that news providers relying on online subscription models should reconsider the current practice of displaying ads even to users who pay for subscriptions. Most subscription-based news websites offer two versions of the website: a free version and a paid version. The free version comes with some restrictions; e.g., non-paying users can access only a subset of the content of the paid version or only a limited number of articles, whereas paying users have unlimited access. However, it is common to expose both sets of users to advertising (Lindsay 2018). Nevertheless, users claim that ads interrupt the web browsing experience, slow it down, and intrude on their privacy (Sołtysik-Piorunkiewicz et al. 2019). Offering a paid but ad-free version of a news website could provide a subscription incentive for loyal users who wish to support the site but do not wish to endure these effects of exposure to ads (Appel et al. 2020; Westcott et al. 2019).

One limitation of our result is the short observation period, especially compared to Shiller et al. (2018), who cover three years and outline a decrease in publishers’ traffic and ad revenue, leading to decreased news production. Further research could investigate how subscription revenue could compensate for the declining ad revenue and how ad blocking causally impacts subscription decisions.

Another limitation of our study is that we could not observe news consumption across multiple news platforms. All the effects we identify are local to our news website. We caution the readers that our results may not generalize to all online news platforms, considering ad blocker's impact differs across websites with different amounts of ads. We could not perform a formal analysis because we do not have information about the specific ads running on our website or its competitors' websites during the data collection period. However, we try to examine the direction of the bias that competing sites might have on our estimated effects by comparing the number of ad slots per page on our website and its competitors' websites, assuming the registered users of our website switched between our and competitor's website. We went to the Internet Archive and used an ad blocker to detect how many blocked ads we find on the historical website of our focal publisher and its competitors' websites[4]. Measured by the number of blocked ad slots, we found that our website displayed 5-12 fewer ads per page than its competitors did. This result indicates that our estimates are more likely to understate the actual effect of ad blocker adoption on overall news consumption. In other words, if there was a bias from users switching websites, then we are likely to have erred on the conservative side. Future research could investigate the effect on aggregate news consumption across websites and how competition between websites plays out.

[4] We define competitors based on Alexa's audience overlap (i.e., websites sharing similar users) and the Top 10 online news brands in the focal country of our news publisher listed by the Digital News Report (Newman et al. 2016).

## Table 1. Description of News Consumption and Users' Behavior on the News Website

### Panel 1. Summary Statistics of News Consumption Variables

| | | Min | Median | Mean | Max | Sd |
|---|---|---|---|---|---|---|
| Main Variables | Article Views | 0.00 | 4.00 | 9.40 | 1204.00 | 17.99 |
| | Breadth | 0.00 | 3.00 | 3.57 | 20.00 | 3.30 |
| Variables across Visits | Visits | 1.00 | 4.00 | 7.72 | 180.00 | 9.88 |
| | Direct Visits | 0.00 | 3.00 | 7.14 | 178.00 | 9.58 |
| | Social Media Visits | 0.00 | 0.00 | 0.12 | 149.00 | 1.37 |
| | Search Engine Visit | 0.00 | 0.00 | 0.55 | 117.00 | 2.70 |
| | Newsletter Visit | 0.00 | 0.00 | 0.00 | 19.00 | 0.10 |
| Variables within Visit | Article Views Per Visit | 0.00 | 1.00 | 1.28 | 213.00 | 2.09 |
| | Other Page Views per Visit | 0.00 | 1.67 | 2.46 | 389.30 | 3.46 |
| | Time per Visit (in seconds) | 0.00 | 259.83 | 408.31 | 36163.00 | 568.76 |
| | Title Length per Visit | 1.00 | 5.50 | 7.39 | 572.00 | 7.13 |
| | Title Length per Article | 1.00 | 4.85 | 4.90 | 97.00 | 1.49 |

### Panel 2. Users' Behavior on the News Website

| Category of News Articles | % of Page Views | Category of Non-News Articles | % of Page Views |
|---|---|---|---|
| International Political News | 8.35 | Homepage | 44.17 |
| Economy News | 6.44 | Account Related Page | 3.72 |
| Sport News | 6.22 | Weather Forecast | 1.82 |
| Regional Political News | 4.79 | Search | 1.02 |
| Finance News | 4.03 | Others | 0.56 |
| Opinion News | 3.81 | Play Page | 0.22 |
| Outlook News | 2.97 | Archive | 0.06 |
| Local Political News | 2.94 | | |
| Art & Culture News | 2.20 | | |
| NewsTicker News | 1.22 | | |
| Sunday News | 1.12 | | |
| Science News | 1.01 | | |
| Digital News | 0.92 | | |
| Lifestyle News | 0.81 | | |
| Photostream News | 0.71 | | |
| Transportation News | 0.31 | | |
| Brief News | 0.30 | | |
| Video News | 0.18 | | |
| Special News | 0.09 | | |
| Data News | 0.02 | | |

Notes: All variables are computed at the week-level.

## Table 2. Before-and-After Analysis of Ad Blocker Adopters and Non-Ad Blocker Adopters

| Variables | Adopters Group Mean | | | Non Adopters Group Mean | | | Diff-in-Diff (DiD) |
|---|---|---|---|---|---|---|---|
| | **Week 12 Adopters** | | | | | | |
| | **Pre-Week 12** | **Post-Week 12** | **Diff** | **Pre-Week 12** | **Post-Week 12** | **Diff** | |
| **Article Views** | 13.50 | 14.40 | 0.90 | 6.11 | 5.11 | -1.00 | 1.90 |
| **Breadth** | 4.87 | 5.34 | 0.47 | 2.77 | 2.52 | -0.25 | 0.72 |
| | **Week 13 Adopters** | | | | | | |
| | **Pre-Week 13** | **Post-Week 13** | **Diff** | **Pre-Week 13** | **Post-Week 13** | **Diff** | |
| **Article Views** | 11.60 | 11.91 | 0.31 | 5.99 | 5.17 | -0.82 | 1.13 |
| **Breadth** | 4.45 | 4.65 | 0.20 | 2.74 | 2.54 | -0.20 | 0.40 |
| | **Week 14 Adopters** | | | | | | |
| | **Pre-Week 14** | **Post-Week 14** | **Diff** | **Pre-Week 14** | **Post-Week 14** | **Diff** | |
| **Article Views** | 10.70 | 11.20 | 0.50 | 5.92 | 5.12 | -0.80 | 1.30 |
| **Breadth** | 4.21 | 4.46 | 0.25 | 2.71 | 2.56 | -0.15 | 0.40 |
| | **Week 15 Adopters** | | | | | | |
| | **Pre-Week 15** | **Post-Week 15** | **Diff** | **Pre-Week 15** | **Post-Week 15** | **Diff** | |
| **Article Views** | 9.27 | 11.60 | 2.33 | 5.85 | 5.04 | -0.81 | 3.14 |
| **Breadth** | 3.93 | 4.42 | 0.49 | 2.69 | 2.67 | -0.02 | 0.51 |
| | **All Adopters** | | | | | | |
| | **Pre-Adoption** | **Post-Adoption** | **Diff** | **Pre-Week 12** | **Post-Week 12** | **Diff** | |
| **Article Views** | 11.46 | 12.35 | 0.89 | 6.11 | 5.11 | -1.00 | 1.89 |
| **Breadth** | 4.41 | 4.78 | 0.37 | 2.77 | 2.52 | -0.25 | 0.62 |
| **Number of Users** | 6,366 | | | 38,290 | | | |

# Table 3. Comparison of Non-Adopters and Adopters Before and After CEM

| Variable | Operationalization | Unmatched Sample | | | Matched Sample | | |
|---|---|---|---|---|---|---|---|
| | | Control Group Mean | Treatment Group Mean | Std. Mean Difference | Control Group Mean | Treatment Group Mean | Std. Mean Difference |
| | **Dummy Variables** | | | | | | |
| **Gender** | **Male=1** | 0.7748 | 0.8144 | 0.1017 | 0.9425 | 0.9425 | 0.0000 |
| **Income** | **Index2=1** | 0.1005 | 0.0861 | -0.0512 | 0.0348 | 0.0348 | 0.0000 |
| | **Index3=1** | 0.1941 | 0.2114 | 0.0424 | 0.1777 | 0.1777 | 0.0000 |
| | **Index4=1** | 0.0849 | 0.0837 | -0.0043 | 0.0314 | 0.0314 | 0.0000 |
| | **Index5=1** | 0.2372 | 0.2273 | -0.0236 | 0.2631 | 0.2631 | 0.0000 |
| | **Inde6=1** | 0.3430 | 0.3411 | -0.0039 | 0.4913 | 0.4913 | 0.0000 |
| **Age** | **25-29=1** | 0.0134 | 0.0135 | 0.0005 | 0.0017 | 0.0017 | 0.0000 |
| | **30-34=1** | 0.0232 | 0.0230 | -0.0013 | 0.0070 | 0.0070 | 0.0000 |
| | **35-39=1** | 0.0458 | 0.0551 | 0.0409 | 0.0279 | 0.0279 | 0.0000 |
| | **40-44=1** | 0.0857 | 0.0956 | 0.0335 | 0.0732 | 0.0732 | 0.0000 |
| | **45-49=1** | 0.1191 | 0.1369 | 0.0515 | 0.1359 | 0.1359 | 0.0000 |
| | **50-54=1** | 0.1333 | 0.1372 | 0.0116 | 0.1794 | 0.1794 | 0.0000 |
| | **55-59=1** | 0.1253 | 0.1131 | -0.0386 | 0.1359 | 0.1359 | 0.0000 |
| | **60-64=1** | 0.1117 | 0.1273 | 0.0470 | 0.1376 | 0.1376 | 0.0000 |
| | **65-69=1** | 0.1116 | 0.0944 | -0.0587 | 0.1220 | 0.1220 | 0.0000 |
| | **70-74=1** | 0.1090 | 0.0912 | -0.0618 | 0.1237 | 0.1237 | 0.0000 |
| | **75-79=1** | 0.0665 | 0.0559 | -0.0462 | 0.0401 | 0.0401 | 0.0000 |
| | **80-85 =1** | 0.0467 | 0.0500 | 0.0153 | 0.0157 | 0.0157 | 0.0000 |
| **First Observed Week** | **Week1=1** | 0.2670 | 0.6022 | 0.4424 | 0.6725 | 0.6725 | 0.0000 |
| | **Week2=1** | 0.1018 | 0.0980 | -0.0129 | 0.0871 | 0.0871 | 0.0000 |
| | **Week3=1** | 0.0819 | 0.0682 | -0.0543 | 0.0540 | 0.0540 | 0.0000 |
| | **Week4=1** | 0.0928 | 0.0504 | -0.1939 | 0.0488 | 0.0488 | 0.0000 |
| | **Week5=1** | 0.1030 | 0.0321 | -0.4018 | 0.0314 | 0.0314 | 0.0000 |
| | **Week6=1** | 0.0771 | 0.0393 | -0.1946 | 0.0401 | 0.0401 | 0.0000 |
| | **Week7=1** | 0.0620 | 0.0290 | -0.1967 | 0.0244 | 0.0244 | 0.0000 |
| | **Week8=1** | 0.0519 | 0.0230 | -0.1928 | 0.0139 | 0.0139 | 0.0000 |
| | **Week9=1** | 0.0343 | 0.0175 | -0.1289 | 0.0087 | 0.0087 | 0.0000 |
| **Last Observed Week** | **Week13=1** | 0.0781 | 0.0210 | -0.3978 | 0.0139 | 0.0139 | 0.0000 |
| | **Week14=1** | 0.1016 | 0.0424 | -0.2932 | 0.0279 | 0.0279 | 0.0000 |
| | **Week15=1** | 0.1595 | 0.0805 | -0.2902 | 0.0575 | 0.0575 | 0.0000 |
| | **Week16=1** | 0.4321 | 0.8497 | 1.1682 | 0.9007 | 0.9007 | 0.0000 |
| **Mode Browser** | **Apple** | 0.2988 | 0.4542 | 0.3119 | 0.5052 | 0.5052 | 0.0000 |
| | **Google** | 0.1726 | 0.1449 | -0.0787 | 0.1080 | 0.1080 | 0.0000 |
| | **Microsoft** | 0.3644 | 0.1869 | -0.4554 | 0.2404 | 0.2404 | 0.0000 |
| | **Mozilla** | 0.1630 | 0.2113 | 0.1183 | 0.1463 | 0.1463 | 0.0000 |
| | **Continuous Variables** | | | | | | |
| **Article Views** | | 3.4593 | 9.7247 | 0.4740 | 4.8999 | 5.1617 | 0.0198 |
| **Breadth** | | 1.6897 | 3.9251 | 0.8995 | 2.6966 | 2.7470 | 0.0202 |
| **Visits** | | 3.5112 | 9.0920 | 0.6784 | 5.2835 | 5.5516 | 0.0326 |
| **Mobile Page Views** | | 1.1944 | 4.0732 | 0.2422 | 1.2444 | 1.3641 | 0.0101 |
| **N** | | 11,665 | 2,499 | | 748 | 574 | |

## Table 4. Treatment Effect on Article Views, Breadth, and Article Views in Each News Category

### Panel 1. Treatment Effect on Article Views and Breadth

| | Ad Blocker Adoption | | Ad Blocker Early Adoption | | Ad Blocker Abandonment | |
|---|---|---|---|---|---|---|
| | **Article Views** | **Breadth** | **Article Views** | **Breadth** | **Article Views** | **Breadth** |
| $\boldsymbol{\beta_1}$ | 0.297*** | 0.198*** | 0.322* | 0.207* | -0.230 | -0.135 |
| | (0.038) | (0.027) | (0.127) | (0.082) | (0.119) | (0.082) |
| $\boldsymbol{\beta_2}$ | 0.155** | 0.096** | -0.010 | 0.019 | 0.109 | 0.067 |
| | (0.049) | (0.036) | (0.133) | (0.086) | (0.154) | (0.101) |
| N | 9,370 | 9,370 | 1,423 | 1,423 | 1,009 | 1,009 |
| $R^2$ | 0.503 | 0.491 | 0.462 | 0.423 | 0.560 | 0.550 |

### Panel 2. Treatment Effect on Article Views in Each News Category

| | Hard News | | | Soft News | | | Non-News Article Pages | | |
|---|---|---|---|---|---|---|---|---|---|
| | **Political** | **Economic** | **Opinion** | **Sports** | **Art & Culture** | **Lifestyle** | **Weather** | **Play Page** | **Account** |
| $\boldsymbol{\beta_1}$ | 0.234*** | 0.145*** | 0.085*** | 0.120*** | 0.061*** | 0.015 | 0.016 | 0.002 | 0.006 |
| | (0.037) | (0.027) | (0.022) | (0.028) | (0.018) | (0.011) | (0.015) | (0.005) | (0.022) |
| $\boldsymbol{\beta_2}$ | 0.125** | 0.088** | 0.056* | 0.034 | 0.030 | 0.013 | 0.012 | 0.001 | 0.000 |
| | (0.046) | (0.031) | (0.027) | (0.030) | (0.022) | (0.014) | (0.018) | (0.008) | (0.030) |
| N | 9,370 | 9,370 | 9,370 | 9,370 | 9,370 | 9,370 | 9,370 | 9,370 | 9,370 |
| $R^2$ | 0.518 | 0.437 | 0.346 | 0.594 | 0.348 | 0.314 | 0.709 | 0.759 | 0.332 |

Notes: In both panels, $\beta_1$ represents the 1-week effect and $\beta_2$ represents the 5-week effect. Each column refers to a separate regression of the following model: $\log(Y_{it} + 1) = \alpha_i + \delta_t + \beta_1 * I_{it1}(within\ 1\ week\ of\ Treatment_{it}) + \beta_2 * I_{it2}(remaining\ weeks\ since\ Treatment_{it}) + \varepsilon_{it}$ on a matched sample centered around 5 weeks (at maximum) before and after treatment starts on week 12. $R^2$ computation includes the explanatory power of the fixed effects. Clustered standard errors appear in parentheses. $^{***}p < 0.001$, $^{**}p < 0.01$, $^{*}p < 0.05$. In Panel 2, $\beta_1$ for Political, Economics, Opinion, and Sports news stay significant ($p < 0.001$) and $\beta_2$ for Political and Economic news stay weakly significant ($p < 0.1$) under the Bonferroni correction of p-values.

**Table 5. Treatment Effects on Within- and Across-Visit Measurements**

**Panel 1. Effects within a Visit (Compatible with a Cognitive Mechanism)**

| | **Article Views per Visits** | **Other Page Views per Visits** | **Title Length per Visit** | **Title Length per Article** | **Time per Visit** |
|---|---|---|---|---|---|
| $\boldsymbol{\beta_1}$ | 0.067*** | 0.047** | 0.146*** | 0.018 | 0.387*** |
| | (0.017) | (0.017) | (0.034) | (0.013) | (0.077) |
| $\boldsymbol{\beta_2}$ | 0.029 | 0.024 | 0.055 | -0.012 | 0.217* |
| | (0.021) | (0.021) | (0.044) | (0.018) | (0.109) |
| N | 9,370 | 9,370 | 9,370 | 8,037 | 9,370 |
| $R^2$ | 0.451 | 0.464 | 0.358 | 0.305 | 0.358 |

**Panel 2. Effect across Visits (Compatible with a Learning Mechanism)**

| | **Visits** | **Direct Visits** | **Social Media Visits** | **Search Engine Visits** | **Newsletter Visits** |
|---|---|---|---|---|---|
| $\boldsymbol{\beta_1}$ | 0.209*** | 0.208*** | -0.004 | 0.039* | 0.001 |
| | (0.025) | (0.027) | (0.008) | (0.017) | (0.001) |
| $\boldsymbol{\beta_2}$ | 0.118*** | 0.112** | 0.008 | 0.036 | -0.000 |
| | (0.033) | (0.035) | (0.012) | (0.021) | (0.001) |
| N | 9,370 | 9,370 | 9,370 | 9,370 | 9,370 |
| $R^2$ | 0.627 | 0.662 | 0.567 | 0.551 | 0.167 |

Notes: $\beta_1$ represents the 1-week effect and $\beta_2$ represents the 5-week effect. Each column refers to a separate regression of the following model: $\log(Y_{it}+1) = \alpha_i + \delta_t + \beta_1 * I_{it1}(within\ 1\ week\ of\ Treatment_{it}) + \beta_2 * I_{it2}(remaining\ weeks\ since\ Treatment_{it}) + \varepsilon_{it}$ on a matched sample centered around 5 weeks (at maximum) before and after treatment starts on week 12. $R^2$ computation includes the explanatory power of the fixed effects. Clustered standard errors appear in parentheses. $^{***}p < 0.001$, $^{**}p < 0.01$, $^{*}p < 0.05$. Under the Bonferroni correction of p-values, Article Views per Visits, Title Length per Visit, Time per Visit,Visits and Direct Visits stay significant ($p < 0.001$).

## Table 6. Heterogeneous Treatment Effects across Users with Different Characteristics

### Panel 1. Single-Browser Users vs. Multiple-Browser Users

| | Single Browser | | Multiple Browsers | |
|---|---|---|---|---|
| | **Article Views** | **Breadth** | **Article Views** | **Breadth** |
| $\boldsymbol{\beta_1}$ | $0.341^{***}$ | $0.219^{***}$ | $0.269^{***}$ | $0.181^{***}$ |
| | (0.065) | (0.046) | (0.048) | (0.034) |
| $\boldsymbol{\beta_2}$ | $0.160^{*}$ | 0.091 | $0.144^{*}$ | 0.090 |
| | (0.076) | (0.054) | (0.064) | (0.047) |
| N | 3,504 | 3,504 | 5,866 | 5,866 |
| $R^2$ | 0.562 | 0.564 | 0.452 | 0.427 |

### Panel 2. Light Users vs. Heavy Users

| | Light Users | | Heavy Users | |
|---|---|---|---|---|
| | **Article Views** | **Breadth** | **Article Views** | **Breadth** |
| $\boldsymbol{\beta_1}$ | $0.333^{***}$ | $0.211^{***}$ | $0.262^{***}$ | $0.183^{***}$ |
| | (0.056) | (0.040) | (0.052) | (0.037) |
| $\boldsymbol{\beta_2}$ | $0.193^{**}$ | $0.135^{**}$ | $0.126^{\cdot}$ | 0.066 |
| | (0.064) | (0.048) | (0.071) | (0.051) |
| N | 4,684 | 4,684 | 4,686 | 4,686 |
| $R^2$ | 0.434 | 0.432 | 0.445 | 0.404 |

Notes: Each column refers to a separate regression of the following model: $\log(Y_{it}+1) = \alpha_i + \delta_t + \beta_1 * I_{it1}(within\ 1\ week\ of\ Treatment_{it}) + \beta_2 * I_{it2}(remaining\ weeks\ since\ Treatment_{it}) + \varepsilon_{it}$ on the matched subsample. Single-browsers users refer to registered users who keep using the same browser to visit our website. Light users refer to registered users who visit our website less than median in the pre-treatment period. $R^2$ computation includes the explanatory power of the fixed effects. Clustered standard errors appear in parentheses. $^{***}p < 0.001$, $^{**}p < 0.01$, $^{*}p < 0.05$.

**Figure 1. Comparison of a News Website without and with an Ad Blocker**

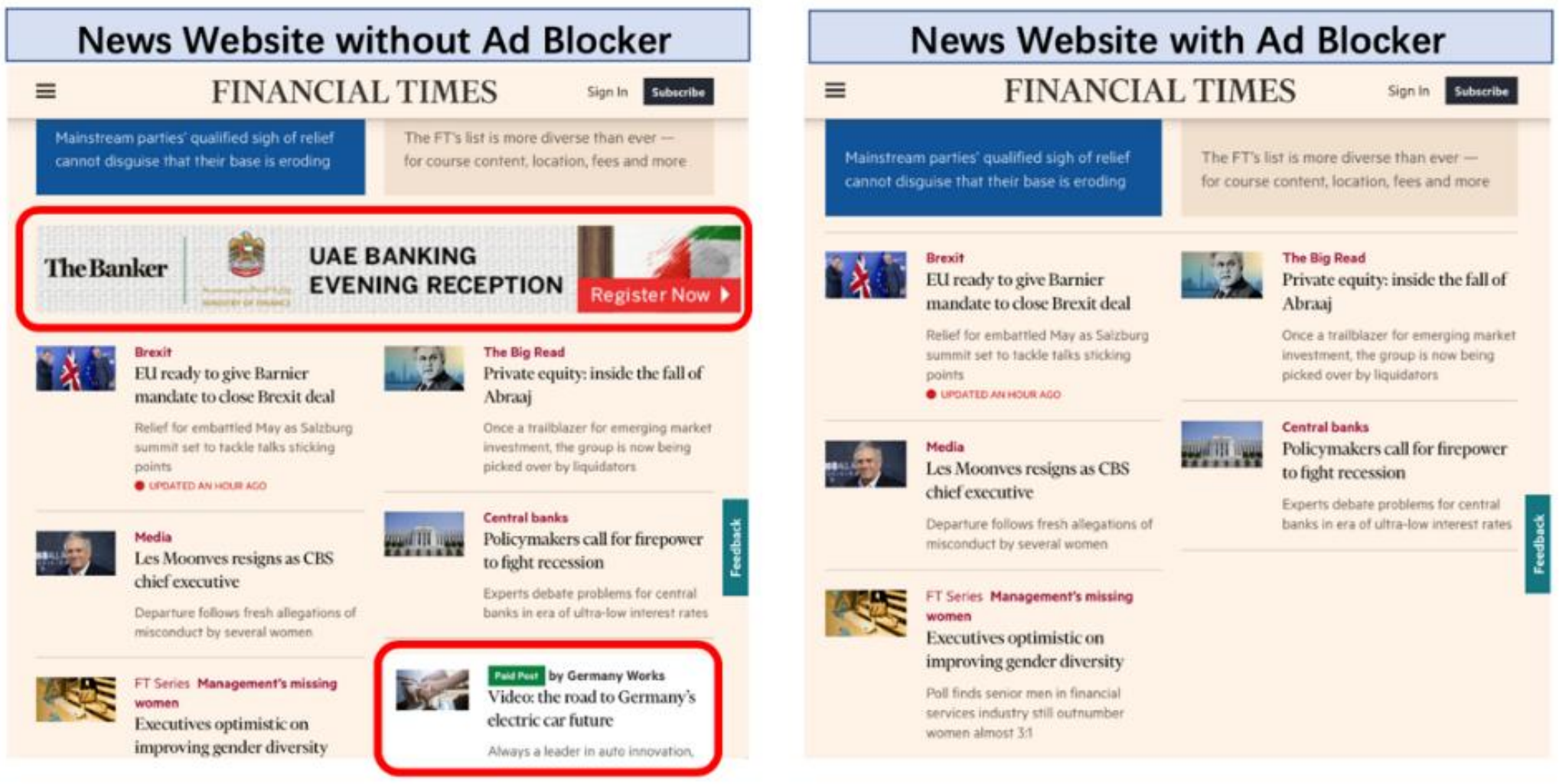

*Source:* Screenshots from *ft.com*

**Figure 2. Construction of Treatment Groups and Control Groups**

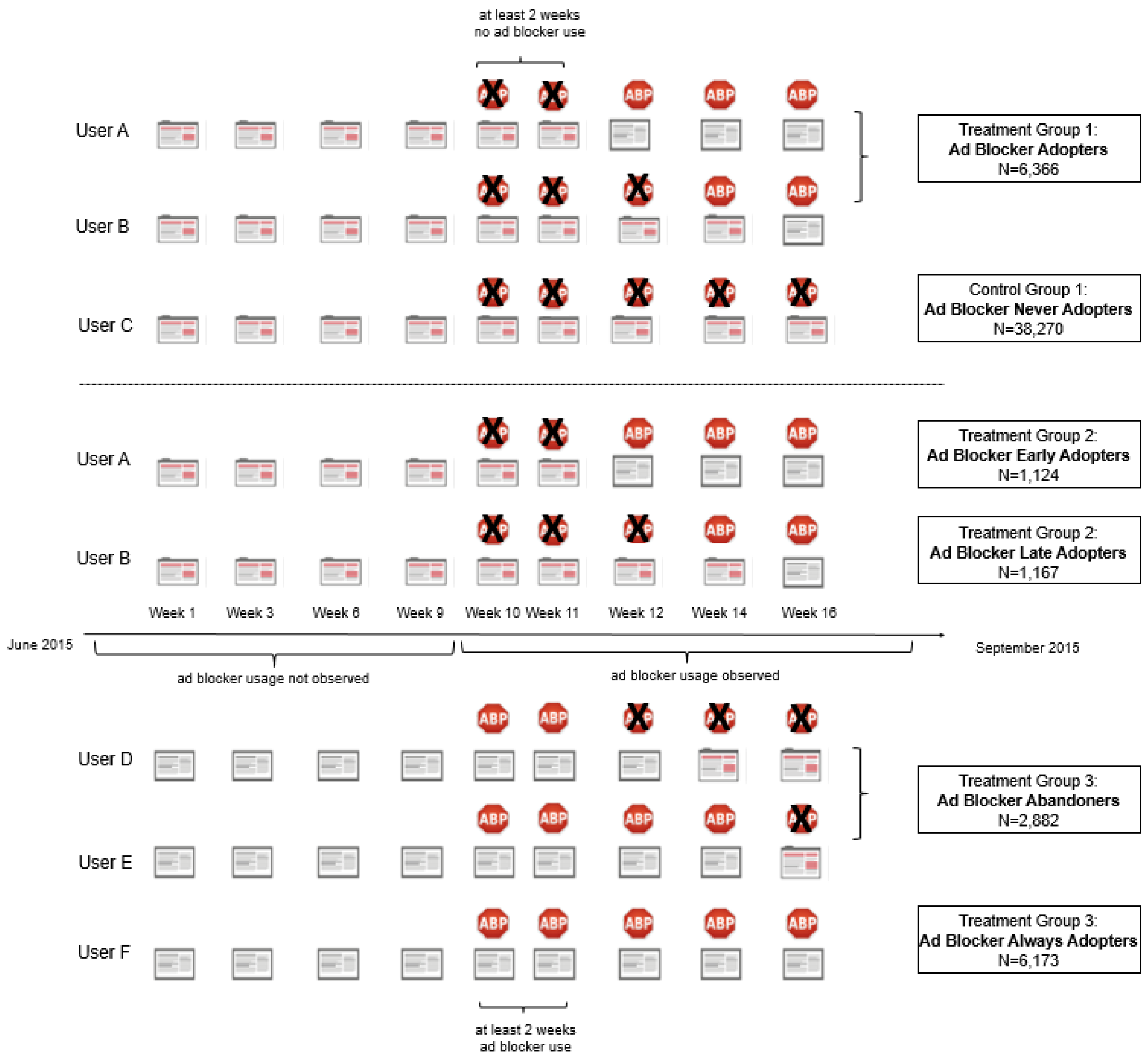

**Figure 3. Distribution of Propensity Score in Matched and Raw Sample**

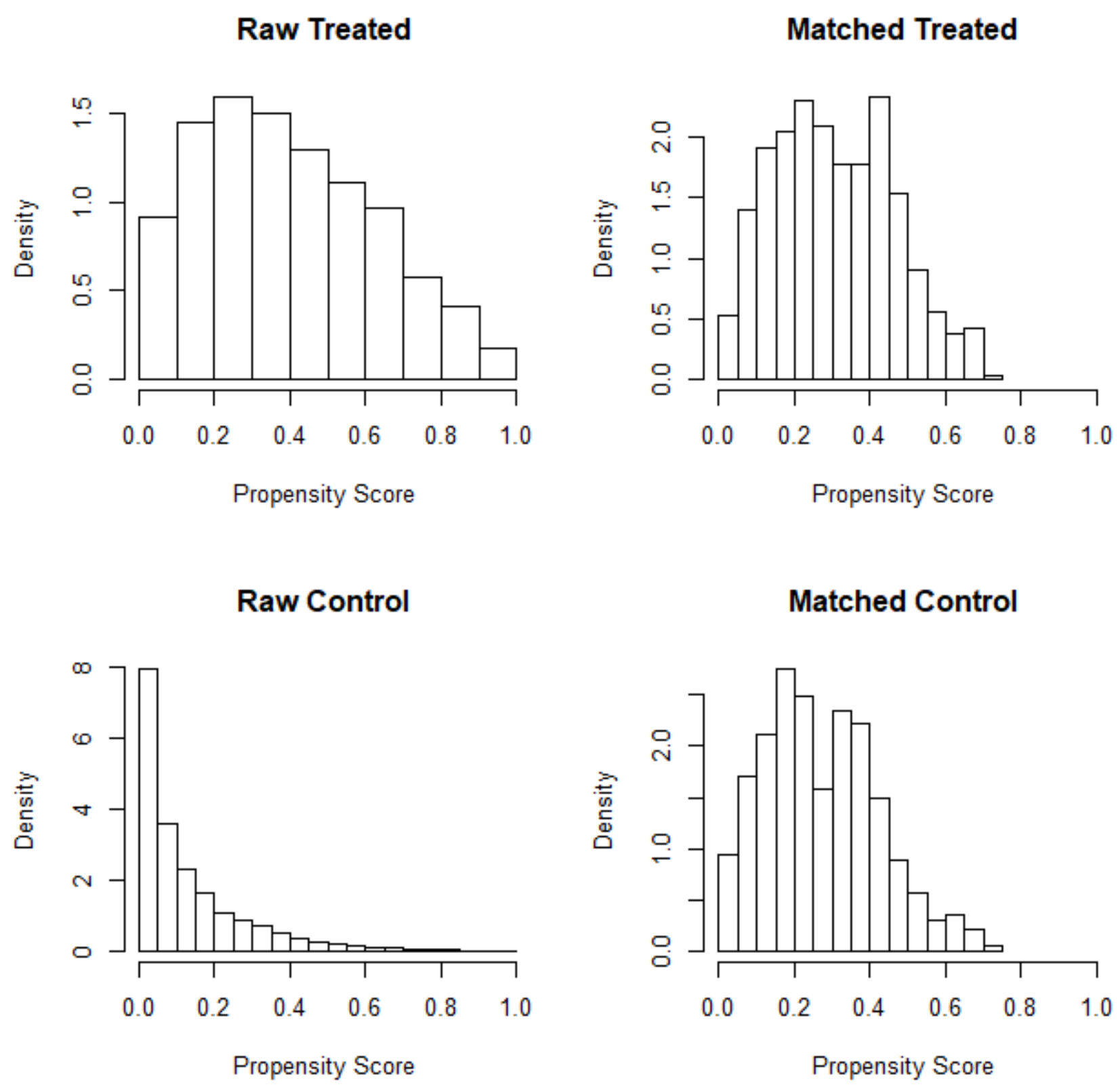

*Web Appendix A. Placebo Treatment Test for Parallel Pre-Treatment Trend*

The identification assumption under difference-in-differences (DiD) is that in the absence of the treatment (i.e., the ad blocker adoption), there would have been no differential changes in the news consumption between the treatment and control group. This assumption also means that we should have comparable changes in the news consumption between both groups before the treatment. The condition fulfilling this assumption is that we have parallel pre-treatment trends. To formally test this parallel pre-treatment trends condition, we perform a “placebo” treatment test by estimating the week-wise treatment effects before and after the treatment. Specifically, we replace the $I_{it1}$ and $I_{it2}$ in equation (1) in the main manuscript for each user *i* with a set of week-wise dummy variables $I_{it-\tau}$ that are equal to 1 if week $t$ is $\tau$ weeks before the treatment and another set of dummy variables $I_{it+\tau}$ that are equal to 1 if week $t$ is $\tau$ weeks after the treatment (and that are equal to zero otherwise):

(2) $$Y_{it} = \alpha_i + \delta_t + \sum_{\tau=2}^{\tau=5} \beta_{-\tau} * I_{it-\tau}(\tau\ weeks\ before\ Treatment_{it}) + \sum_{\tau=0}^{\tau=4} \beta_{\tau} * I_{it+\tau}(\tau\ weeks\ since\ Treatment_{it}) + \varepsilon_{it},$$

where $Y_{it}$ is the news consumption for user *i* in week $t$; $\alpha_i$ is a user-level fixed effect; $\delta_t$ is the week-fixed effect; $\varepsilon_{it}$ is the standard error clustered at the user-level. We choose the last week before treatment ($I_{it-1}$) as the omitted default category. If the trends of the treatment and control group are parallel, then the parameters $\beta_{-\tau}$ will be statistically indistinguishable from zero. As reported in Table S1, all the main news consumption measures we use pass this test because the parameters $\beta_{-\tau}$ in the pre-treatment period have non-significant point estimates.

# Table S1. Placebo Treatment Test on News Consumption Variables

| | **Article Views** | **Breadth** | **Visits** | **Article Views per Visit** | **International Political News** | **Regional Political News** | **Local Political News** | **Economy News** |
|---|---|---|---|---|---|---|---|---|
| $\beta_{-2}$ | -0.054 | -0.065 | -0.023 | -0.029 | 0.007 | -0.029 | -0.012 | -0.003 |
| | (0.042) | (0.031) | (0.028) | (0.022) | (0.030) | (0.030) | (0.027) | (0.027) |
| $\beta_{-3}$ | -0.010 | -0.037 | 0.037 | -0.023 | 0.054 | -0.007 | 0.006 | -0.016 |
| | (0.047) | (0.035) | (0.030) | (0.023) | (0.034) | (0.031) | (0.028) | (0.031) |
| $\beta_{-4}$ | -0.014 | -0.039 | 0.018 | -0.013 | -0.018 | -0.004 | 0.005 | -0.010 |
| | (0.049) | (0.036) | (0.032) | (0.024) | (0.032) | (0.032) | (0.027) | (0.033) |
| $\beta_{-5}$ | -0.048 | -0.054 | -0.039 | 0.004 | -0.015 | -0.060 | -0.016 | -0.038 |
| | (0.047) | (0.035) | (0.035) | (0.024) | (0.032) | (0.033) | (0.027) | (0.030) |
| $R^2$ | 0.505 | 0.492 | 0.628 | 0.452 | 0.487 | 0.455 | 0.426 | 0.438 |
| N | 9,370 | 9,370 | 9,370 | 9,370 | 9,370 | 9,370 | 9,370 | 9,370 |

| | **Finance News** | **Opinion News** | **Sport News** | **Art & Culture News** | **Lifestyle News** | **Weather Forecast** | **Play Page** | **Account** |
|---|---|---|---|---|---|---|---|---|
| $\beta_{-2}$ | -0.016 | -0.001 | -0.013 | -0.009 | 0.001 | 0.020 | -0.004 | 0.011 |
| | (0.029) | (0.027) | (0.029) | (0.018) | (0.015) | (0.017) | (0.004) | (0.027) |
| $\beta_{-3}$ | -0.013 | -0.007 | 0.064 | -0.017 | 0.019 | 0.007 | -0.000 | 0.022 |
| | (0.030) | (0.029) | (0.030) | (0.021) | (0.018) | (0.022) | (0.004) | (0.028) |
| $\beta_{-4}$ | -0.030 | -0.002 | 0.019 | 0.001 | -0.023 | -0.023 | -0.001 | 0.035 |
| | (0.029) | (0.030) | (0.029) | (0.021) | (0.017) | (0.021) | (0.004) | (0.032) |
| $\beta_{-5}$ | -0.064 | -0.004 | 0.002 | 0.005 | -0.028 | -0.002 | -0.002 | -0.021 |
| | (0.029) | (0.030) | (0.030) | (0.022) | (0.014) | (0.020) | (0.005) | (0.027) |
| $R^2$ | 0.575 | 0.347 | 0.595 | 0.348 | 0.315 | 0.709 | 0.759 | 0.332 |
| N | 9,370 | 9,370 | 9,370 | 9,370 | 9,370 | 9,370 | 9,370 | 9,370 |

| | **Home Page Views per Visit** | **Words per Article** | **Words per Sentence** | **Time per Visit** | **Direct Visits** | **Social Media Visits** | **Search Engine Visits** | **Newsletter Visits** |
|---|---|---|---|---|---|---|---|---|
| $\beta_{-2}$ | -0.018 | -0.061 | -0.004 | -0.121 | -0.029 | -0.002 | 0.016 | 0.000 |
| | (0.017) | (0.051) | (0.008) | (0.110) | (0.030) | (0.008) | (0.021) | (0.000) |
| $\beta_{-3}$ | -0.004 | -0.011 | -0.003 | -0.044 | 0.042 | 0.010 | 0.008 | 0.001 |
| | (0.017) | (0.052) | (0.009) | (0.114) | (0.031) | (0.009) | (0.023) | (0.001) |
| $\beta_{-4}$ | 0.029 | -0.045 | -0.004 | 0.107 | 0.026 | 0.003 | -0.013 | 0.001 |
| | (0.019) | (0.056) | (0.008) | (0.116) | (0.034) | (0.009) | (0.023) | (0.000) |
| $\beta_{-5}$ | 0.012 | 0.013 | -0.009 | -0.023 | -0.022 | 0.004 | -0.040 | 0.000 |
| | (0.019) | (0.049) | (0.010) | (0.119) | (0.036) | (0.007) | (0.021) | (0.000) |
| $R^2$ | 0.514 | 0.364 | 0.248 | 0.360 | 0.662 | 0.568 | 0.551 | 0.167 |
| N | 9,370 | 9,370 | 9,370 | 9,370 | 9,370 | 9,370 | 9,370 | 9,370 |

Notes: Each column refers to a separate regression with the following model: $\log(Y_{it} + 1) = \alpha_i + \delta_t + \sum_{\tau=2}^{\tau=5} \beta_{-\tau} * I_{it-\tau}(\tau\ weeks\ before\ Treatment_{it}) + \sum_{\tau=0}^{\tau=4} \beta_\tau * I_{it+\tau}(\tau\ weeks\ since\ Treatment_{it}) + \varepsilon_{it}$ on the matched sample of ad blocker adopters and non-adopters. $\beta_{-\tau}$ are the placebo treatment effects and are reported with $\beta_{-1}$ omitted as the default category. $R^2$ computation includes the explanatory power of the fixed effects. Standard errors clustered at the user-level appear in parentheses. $^{*}p < 0.01$.

*Web Appendix B. Robustness Check on Adding Time-Varying Controls*

The placebo treatment test (reported in Table S1) statistically validates the identification condition (parallel pre-treatment trend) of DiD. Recall that DiD removes all time-invariant confounders. If the parallel pre-treatment trend holds, then DiD will also eliminate any bias from time-varying confounders. The reason is that a common pre-treatment trend implies that time-varying confounders, if any, impact both groups (i.e., treatment and control group) in the same way in the pre-treatment period.

Concerns may remain that a time-varying confounder kicks in at the same time when the treatment occurs and, thus, will bias our result. For example, a user reads news on a browser with no ad-blocking feature. Then, the user installs an additional more user-friendly browser with an ad-blocking feature and, at the same time, starts reading the news with multiple browsers. So, this installation of a browser changes her ad blocker usage, and the more user-friendly browser also impacts her news reading behavior.

To further establish the robustness of our main result, we rerun our main estimation by adding the following time-varying control variables: browser switching (i.e., the number of different browsers that a user uses during a particular week), ordering (i.e., the number of orders that a user places on the website during a particular week, such as purchasing access to the news archive), and commenting (i.e., the number of comments that a user leaves during a particular week).

Specifically, we estimate the following model

(3)
$$Y_{it} = \alpha_i + \delta_t + \beta_1 * I_{it1}(within\ 1\ week\ of\ Treatment_{it})$$
$$+\beta_2 * I_{it2}(remaining\ weeks\ since\ Treatment_{it}) + \beta_3 * Browsers_{it} + \beta_4 * Orders_{it}$$
$$+ \beta_5 * Comments_{it} + \varepsilon_{it}$$

Table S2 reports the result that our main treatment effects ($\beta_1$ and $\beta_2$) remain very robust because they are similar to the treatment effects reported in Table 4. For brevity, we continue and only report results for news categories classified into the following: hard news (political, economic and opinion news) and soft news (sports, culture & art, lifestyle news).

## Table S2. Robustness Check of Main Model after also Controlling for Browser Switching, Ordering, & Commenting

| | Article Views | Breadth | Visits | Article Views per Visit | Hard News | Soft News |
|---|---|---|---|---|---|---|
| $\beta_1$ | 0.255*** | 0.170*** | 0.169*** | 0.065*** | 0.244*** | 0.133*** |
| | (0.037) | (0.026) | (0.024) | (0.017) | (0.038) | (0.031) |
| $\beta_2$ | 0.119* | 0.071* | 0.086** | 0.026 | 0.143** | 0.028 |
| | (0.047) | (0.035) | (0.032) | (0.021) | (0.048) | (0.035) |
| $\beta_3$ $(Browsers_{it})$ | 0.373*** | 0.250*** | 0.342*** | 0.024* | 0.335*** | 0.161*** |
| | (0.023) | (0.018) | (0.016) | (0.010) | (0.023) | (0.019) |
| $\beta_4$ $(Orders_{it})$ | -0.291 | -0.352* | -0.040 | -0.218 | -0.208 | -0.115 |
| | (0.286) | (0.150) | (0.151) | (0.130) | (0.239) | (0.143) |
| $\beta_5(Comments_{it})$ | -0.035 | -0.012 | 0.032 | -0.049** | -0.003 | 0.028 |
| | (0.032) | (0.024) | (0.017) | (0.016) | (0.034) | (0.030) |
| N | 9,370 | 9,370 | 9,370 | 9,370 | 9,370 | 9,370 |
| $R^2$ | 0.529 | 0.514 | 0.663 | 0.453 | 0.558 | 0.548 |

Notes: $\beta_1$ represents the 1-week effect and $\beta_2$ represents the 5-week effect. ***$p < 0.001$, **$p < 0.01$, *$p < 0.05$. Each column refers to a separate regression of the following model on the matched sample of ad blocker adopters and non-adopters: $\log(Y_{it} + 1) = \alpha_i + \delta_t + \beta_1 * I_{it1}(within\ 1\ week\ of\ Treatment_{it}) + \beta_2 * I_{it2}(remaining\ weeks\ since\ Treatment_{it}) + \beta_3 * Browsers_{it} + \beta_4 * Orders_{it} + \beta_5 * Comments_{it} + \varepsilon_{it}$. $R^2$ computation includes the explanatory power of the fixed effects. Standard errors clustered at the user-level appear in parentheses.

*Web Appendix C. Robustness Checks on Matching Method*

In this section, we check for the robustness of the results of our matching method and, in particular, of the coarsened exact matching (CEM) method.

First, we check for the robustness of our result by running specification (1) from the main manuscript using the unmatched sample. The results are reported in Table S3 and remain similar.

Second, we run specification (4) to check for the robustness of missing observations due to matching on demographics (i.e., age, gender, income). The demographics data came from the customer relationship dataset (CRM) of the news publisher. For the demographic values that are missing, which could result from a user deliberately not reporting that value or a coding error, we add the interaction (i.e., $I_{it2}$ in specification (4)) between the missing value and our main effect. We report the results in Table S4. The coefficients ($\beta_3$ $and$ $\beta_4$) of the interaction terms are statistically insignificant, which indicates that these missing values occur at random (and, therefore, do not induce bias in the estimation). In addition, our main effect remains similar.

(4)
$$Y_{it} = \alpha_i + \delta_t + \beta_1 * I_{it1}(within\ 1\ week\ of\ Treatment_{it}) + \beta_2 * I_{it2}(remaining\ weeks\ since\ Treatment_{it}) + \beta_3 * I_{it1}(within\ 1\ week\ of\ Treatment_{it}) * Missing_i + \beta_4 * I_{it2}(remaining\ weeks\ since\ Treatment_{it}) * Missing_i + \varepsilon_{it}.$$

Third, we check for the robustness of controlling for observed variables (rather than using them in the matching; see Table 3 or Table S5 for the list of these variables). In addition, we add an interaction between the week-fixed effect and the browser mode in the pre-treatment period because we know that the Apple iOS 9 event triggered ad blocker adoption and, thus, users with different browsers might experience different time trends. We run the specification (5) and report the results in Table S5. Our main effects again remain similar.

(5)

$$Y_{it} = \beta_1 * I_{it1}(within\ 1\ week\ of\ Treatment_{it})$$
$$+\beta_2 * I_{it2}(remaining\ weeks\ since\ Treatment_{it}) + \sum_{\tau=1}^{\tau=n} \gamma_\tau *$$
$$MatchingVariables_{i\tau} + \delta_t$$
$$+ \eta_{it} * (\delta_t * ModeBrowser_i) + \varepsilon_{it}$$

Forth, we check for the robustness of the results of coarsened exact matching. We rerun specification (1) of the main manuscript on the sample matched using propensity score matching. The results (reported in Table S6) continue to remain similar.

**Table S3. Robustness Check on Unmatched Sample**

| | Article Views | Breadth | Visits | Article Views per Visit | Hard News | Soft News |
|---|---|---|---|---|---|---|
| $\beta_1$ | 0.260*** | 0.159*** | 0.231*** | 0.032*** | 0.229*** | 0.185*** |
| | (0.009) | (0.006) | (0.007) | (0.004) | (0.010) | (0.008) |
| $\beta_2$ | 0.110*** | 0.067*** | 0.122*** | 0.002 | 0.120*** | 0.122*** |
| | (0.012) | (0.007) | (0.008) | (0.005) | (0.012) | (0.009) |
| N | 252,428 | 252,428 | 252,428 | 252,428 | 252,428 | 252,428 |
| $R^2$ | 0.631 | 0.624 | 0.685 | 0.541 | 0.633 | 0.606 |

Notes: $\beta_1$ represents the 1-week effect and $\beta_2$ represents the 5-week effect. Each column refers to a separate regression of the following model on the unmatched sample: $\log(Y_{it} + 1) = \alpha_i + \delta_t + \beta_1 * I_{it1}(within\ 1\ week\ of\ Treatment_{it}) + \beta_2 * I_{it2}(remaining\ weeks\ since\ Treatment_{it}) + \varepsilon_{it}$. $R^2$ computation includes the explanatory power of the fixed effects. Standard errors clustered at the user-level appear in parentheses. ***p < 0.001, **p < 0.01, *p < 0.05.

**Table S4. Robustness Check on Missing Observations in Demographics**

| | Article Views | Breadth | Visits | Article Views per Visit | Hard News | Soft News |
|---|---|---|---|---|---|---|
| $\beta_1$ | 0.245*** | 0.151*** | 0.212*** | 0.036*** | 0.216*** | 0.198*** |
| | (0.015) | (0.010) | (0.010) | (0.007) | (0.015) | (0.013) |
| $\beta_2$ | 0.088*** | 0.052*** | 0.096*** | 0.007 | 0.119*** | 0.115*** |
| | (0.017) | (0.012) | (0.013) | (0.008) | (0.018) | (0.015) |
| $\beta_3$ | 0.010 | -0.008 | 0.006 | -0.001 | 0.023 | -0.031 |
| | (0.022) | (0.014) | (0.015) | (0.010) | (0.022) | (0.019) |
| $\beta_4$ | 0.026 | 0.001 | 0.027 | -0.002 | 0.012 | 0.007 |
| | (0.025) | (0.016) | (0.018) | (0.010) | (0.026) | (0.021) |
| N | 118,696 | 118,696 | 118,696 | 118,696 | 118,696 | 118,696 |
| $R^2$ | 0.679 | 0.668 | 0.745 | 0.562 | 0.676 | 0.636 |

Notes: Each column refers to a separate regression of the following model on the unmatched sample: $\log(Y_{it} + 1) = \alpha_i + \delta_t + \beta_1 * I_{it1}(within\ 1\ week\ of\ Treatment_{it}) + \beta_2 * I_{it2}(remaining\ weeks\ since\ Treatment_{it}) + \beta_3 * I_{it1}(within\ 1\ week\ of\ Treatment_{it}) * Missing_i + \beta_4 * I_{it2}(remaining\ weeks\ since\ Treatment_{it} for\ treated\ * Missing_i) + \varepsilon_{it}$. $\beta_1$ represents the 5-week effect, and $\beta_2$ represents the interaction effect of the 5-week effect and any missing observations (e.g., due to users not revealing full information in our CRM data). Insignificant $\beta_2$ indicates matching does not induce bias in the estimation. $R^2$ computation includes the explanatory power of the fixed effects. Standard errors clustered at the user-level appear in parentheses. ***p < 0.001, **p < 0.01, *p < 0.05.

## Table S5. Robustness Check on Controlling for Observed Variables (Instead of Using Matching Method)

| Independent Variables \ Dependent Variables | Article Views | Breadth | Visits | Article Views per Visit | Hard News | Soft News |
|---|---|---|---|---|---|---|
| (Intercept) | -0.277 | -0.148 | 0.481** | 0.091 | -0.224 | -0.552* |
| | (0.226) | (0.161) | (0.159) | (0.129) | (0.231) | (0.215) |
| $I_{it1}$ | 0.302*** | 0.196*** | 0.290*** | 0.042*** | 0.227*** | 0.206*** |
| | (0.013) | (0.009) | (0.009) | (0.007) | (0.013) | (0.012) |
| $I_{it2}$ | 0.175*** | 0.127*** | 0.239*** | -0.004 | 0.157*** | 0.140*** |
| | (0.014) | (0.010) | (0.010) | (0.008) | (0.014) | (0.013) |
| Gender | -0.007 | 0.009 | 0.016** | -0.016*** | -0.021** | -0.004 |
| | (0.007) | (0.005) | (0.005) | (0.004) | (0.007) | (0.007) |
| Income2 | 0.015 | -0.018 | -0.007 | 0.006 | -0.012 | 0.034* |
| | (0.017) | (0.012) | (0.012) | (0.010) | (0.017) | (0.016) |
| Income3 | 0.040* | 0.006 | 0.011 | 0.007 | 0.014 | 0.027 |
| | (0.016) | (0.011) | (0.011) | (0.009) | (0.016) | (0.015) |
| Income4 | 0.006 | -0.010 | 0.002 | -0.009 | 0.003 | -0.005 |
| | (0.017) | (0.012) | (0.012) | (0.010) | (0.018) | (0.016) |
| Income5 | 0.018 | -0.003 | 0.010 | -0.002 | -0.008 | 0.013 |
| | (0.016) | (0.011) | (0.011) | (0.009) | (0.016) | (0.015) |
| Income6 | 0.028 | 0.000 | 0.004 | 0.004 | 0.010 | -0.007 |
| | (0.015) | (0.011) | (0.011) | (0.009) | (0.016) | (0.015) |
| Age25-29 | 0.051 | 0.045 | 0.024 | 0.018 | 0.070 | 0.030 |
| | (0.041) | (0.029) | (0.029) | (0.023) | (0.042) | (0.039) |
| Age30-34 | 0.009 | -0.004 | 0.013 | -0.012 | 0.021 | -0.038 |
| | (0.038) | (0.027) | (0.027) | (0.022) | (0.039) | (0.036) |
| Age35-39 | 0.015 | 0.030 | 0.020 | -0.019 | 0.045 | -0.103** |
| | (0.035) | (0.025) | (0.025) | (0.020) | (0.036) | (0.033) |
| Age40-44 | 0.008 | 0.014 | 0.030 | -0.034 | -0.000 | -0.010 |
| | (0.034) | (0.024) | (0.024) | (0.019) | (0.035) | (0.032) |
| Age45-49 | -0.014 | 0.016 | 0.028 | -0.031 | -0.017 | 0.001 |
| | (0.033) | (0.024) | (0.023) | (0.019) | (0.034) | (0.032) |
| Age50-54 | -0.033 | -0.006 | 0.006 | -0.033 | -0.027 | -0.010 |
| | (0.033) | (0.024) | (0.023) | (0.019) | (0.034) | (0.032) |
| Age55-59 | -0.014 | 0.008 | 0.013 | -0.022 | -0.014 | 0.034 |
| | (0.033) | (0.024) | (0.024) | (0.019) | (0.034) | (0.032) |
| Age60-64 | -0.039 | -0.012 | 0.003 | -0.035 | -0.017 | 0.020 |
| | (0.034) | (0.024) | (0.024) | (0.019) | (0.034) | (0.032) |
| Age65-69 | -0.025 | -0.012 | -0.008 | -0.020 | 0.002 | 0.033 |
| | (0.034) | (0.024) | (0.024) | (0.019) | (0.034) | (0.032) |
| Age70-74 | -0.028 | -0.010 | -0.009 | -0.023 | 0.020 | 0.001 |
| | (0.034) | (0.024) | (0.024) | (0.019) | (0.034) | (0.032) |
| Age75-79 | 0.003 | 0.006 | 0.020 | -0.021 | 0.029 | 0.038 |
| | (0.034) | (0.024) | (0.024) | (0.020) | (0.035) | (0.033) |
| Age80-84 | 0.034 | 0.031 | 0.048 | -0.023 | 0.079* | 0.026 |
| | (0.035) | (0.025) | (0.025) | (0.020) | (0.036) | (0.033) |
| Pre-ArticleViews | 0.006*** | -0.005*** | -0.008*** | 0.008*** | 0.004*** | 0.004*** |
| | (0.000) | (0.000) | (0.000) | (0.000) | (0.000) | (0.000) |
| Pre-Breadth | 0.254*** | 0.218*** | 0.053*** | 0.121*** | 0.245*** | 0.113*** |

| Independent Variables \ Dependent Variables | Article Views | Breadth | Visits | Article Views per Visit | Hard News | Soft News |
|---|---|---|---|---|---|---|
| | (0.002) | (0.001) | (0.001) | (0.001) | (0.002) | (0.002) |
| Pre-Visits | 0.001$^{*}$ | -0.002$^{***}$ | 0.069$^{***}$ | -0.036$^{***}$ | 0.000 | 0.005$^{***}$ |
| | (0.001) | (0.000) | (0.000) | (0.000) | (0.001) | (0.001) |
| Firstweek1 | 0.172$^{***}$ | 0.133$^{***}$ | 0.090$^{***}$ | 0.083$^{***}$ | 0.085$^{***}$ | 0.045$^{*}$ |
| | (0.020) | (0.014) | (0.014) | (0.011) | (0.020) | (0.019) |
| Firstweek2 | 0.122$^{***}$ | 0.095$^{***}$ | 0.018 | 0.097$^{***}$ | 0.058$^{**}$ | -0.018 |
| | (0.021) | (0.015) | (0.015) | (0.012) | (0.021) | (0.020) |
| Firstweek3 | 0.072$^{***}$ | 0.065$^{***}$ | -0.011 | 0.088$^{***}$ | 0.018 | -0.041$^{*}$ |
| | (0.021) | (0.015) | (0.015) | (0.012) | (0.022) | (0.020) |
| Firstweek4 | 0.055$^{*}$ | 0.041$^{**}$ | -0.029 | 0.076$^{***}$ | 0.018 | -0.027 |
| | (0.022) | (0.015) | (0.015) | (0.012) | (0.022) | (0.020) |
| Firstweek5 | 0.041 | 0.024 | -0.022 | 0.062$^{***}$ | 0.004 | -0.020 |
| | (0.022) | (0.015) | (0.015) | (0.012) | (0.022) | (0.021) |
| Firstweek6 | 0.079$^{***}$ | 0.059$^{***}$ | -0.027 | 0.097$^{***}$ | 0.052$^{*}$ | -0.009 |
| | (0.022) | (0.016) | (0.016) | (0.013) | (0.023) | (0.021) |
| Firstweek7 | 0.028 | 0.012 | -0.023 | 0.057$^{***}$ | 0.015 | -0.012 |
| | (0.022) | (0.016) | (0.016) | (0.013) | (0.023) | (0.021) |
| Firstweek8 | 0.038 | 0.033$^{*}$ | -0.011 | 0.056$^{***}$ | 0.009 | -0.016 |
| | (0.023) | (0.017) | (0.016) | (0.013) | (0.024) | (0.022) |
| Firstweek9 | 0.031 | 0.014 | -0.051$^{**}$ | 0.067$^{***}$ | -0.016 | -0.004 |
| | (0.027) | (0.019) | (0.019) | (0.015) | (0.027) | (0.025) |
| Firstweek10 | 0.053$^{*}$ | 0.033 | -0.024 | 0.052$^{***}$ | 0.040 | -0.009 |
| | (0.026) | (0.019) | (0.018) | (0.015) | (0.027) | (0.025) |
| Lastweek11 | 0.060$^{*}$ | 0.034 | -0.003 | 0.041$^{*}$ | 0.033 | 0.017 |
| | (0.028) | (0.020) | (0.020) | (0.016) | (0.029) | (0.027) |
| Lastweek12 | 0.058$^{*}$ | 0.043$^{*}$ | -0.009 | 0.058$^{***}$ | 0.018 | 0.012 |
| | (0.027) | (0.020) | (0.019) | (0.016) | (0.028) | (0.026) |
| Lastweek13 | 0.090$^{***}$ | 0.074$^{***}$ | 0.007 | 0.077$^{***}$ | 0.031 | 0.017 |
| | (0.026) | (0.018) | (0.018) | (0.015) | (0.026) | (0.024) |
| Lastweek14 | 0.123$^{***}$ | 0.100$^{***}$ | 0.015 | 0.105$^{***}$ | 0.040 | 0.029 |
| | (0.024) | (0.017) | (0.017) | (0.014) | (0.025) | (0.023) |
| Lastweek15 | 0.172$^{***}$ | 0.150$^{***}$ | 0.085$^{***}$ | 0.114$^{***}$ | 0.055$^{*}$ | 0.031 |
| | (0.023) | (0.016) | (0.016) | (0.013) | (0.023) | (0.022) |
| Lastweek16 | 0.330$^{***}$ | 0.260$^{***}$ | 0.288$^{***}$ | 0.119$^{***}$ | 0.182$^{***}$ | 0.066$^{**}$ |
| | (0.022) | (0.016) | (0.015) | (0.012) | (0.022) | (0.021) |
| Pre-Mobile Page Views | 0.011$^{***}$ | 0.005$^{***}$ | 0.007$^{***}$ | 0.002$^{***}$ | 0.011$^{***}$ | 0.009$^{***}$ |
| | (0.000) | (0.000) | (0.000) | (0.000) | (0.000) | (0.000) |
| Week8 | 0.195 | 0.106 | -0.061 | 0.216 | 0.112 | -0.117 |
| | (0.342) | (0.243) | (0.241) | (0.195) | (0.349) | (0.325) |
| Week9 | 0.051 | 0.199 | -0.096 | 0.181 | 0.091 | -0.021 |
| | (0.342) | (0.243) | (0.240) | (0.195) | (0.349) | (0.325) |
| Week10 | 0.445 | 0.377 | 0.129 | 0.283 | 0.133 | 0.342 |
| | (0.302) | (0.214) | (0.212) | (0.172) | (0.308) | (0.286) |
| Week11 | 0.429 | 0.368 | 0.091 | 0.251 | 0.324 | 0.314 |
| | (0.308) | (0.218) | (0.216) | (0.175) | (0.314) | (0.292) |
| Week12 | 0.435 | 0.469$^{*}$ | 0.292 | 0.164 | 0.267 | 0.277 |
| | (0.322) | (0.229) | (0.226) | (0.183) | (0.328) | (0.305) |
| Week13 | 0.838$^{*}$ | 0.513$^{*}$ | 0.181 | 0.487$^{**}$ | 0.735$^{*}$ | 0.334 |
| | (0.331) | (0.235) | (0.233) | (0.188) | (0.338) | (0.314) |

| Independent Variables \ Dependent Variables | **Article Views** | **Breadth** | **Visits** | **Article Views per Visit** | **Hard News** | **Soft News** |
|---|---|---|---|---|---|---|
| Week14 | 0.576 | 0.322 | 0.243 | 0.203 | 0.466 | 0.479 |
| | (0.314) | (0.223) | (0.221) | (0.179) | (0.321) | (0.298) |
| Week15 | 0.399 | 0.230 | -0.093 | 0.397* | 0.378 | 0.233 |
| | (0.322) | (0.229) | (0.226) | (0.183) | (0.329) | (0.305) |
| Week16 | 0.329 | 0.023 | -0.078 | 0.280 | 0.525 | 0.180 |
| | (0.322) | (0.229) | (0.226) | (0.183) | (0.329) | (0.305) |
| ModeBrowserApple | 0.429 | 0.331* | 0.238 | 0.179 | 0.263 | 0.537* |
| | (0.223) | (0.158) | (0.157) | (0.127) | (0.227) | (0.211) |
| ModeBrowserGoogle | 0.338 | 0.263 | 0.243 | 0.110 | 0.236 | 0.418* |
| | (0.224) | (0.159) | (0.157) | (0.127) | (0.228) | (0.212) |
| ModeBrowserMicrosoft | 0.493* | 0.335* | 0.315* | 0.183 | 0.329 | 0.487* |
| | (0.223) | (0.158) | (0.157) | (0.127) | (0.227) | (0.211) |
| ModeBrowserMozilla | 0.439* | 0.318* | 0.324* | 0.124 | 0.300 | 0.526* |
| | (0.224) | (0.159) | (0.157) | (0.127) | (0.228) | (0.212) |
| Week8*ModeBrowserApple | -0.149 | -0.069 | 0.066 | -0.194 | -0.038 | 0.118 |
| | (0.343) | (0.244) | (0.241) | (0.195) | (0.350) | (0.325) |
| Week9*ModeBrowserApple | -0.045 | -0.203 | 0.057 | -0.156 | -0.066 | 0.012 |
| | (0.343) | (0.244) | (0.241) | (0.195) | (0.350) | (0.325) |
| Week10*ModeBrowserApple | -0.428 | -0.360 | -0.166 | -0.256 | -0.152 | -0.297 |
| | (0.303) | (0.215) | (0.213) | (0.172) | (0.309) | (0.287) |
| Week11*ModeBrowserApple | -0.451 | -0.372 | -0.128 | -0.246 | -0.374 | -0.297 |
| | (0.308) | (0.219) | (0.217) | (0.175) | (0.315) | (0.292) |
| Week12*ModeBrowserApple | -0.396 | -0.434 | -0.291 | -0.150 | -0.252 | -0.243 |
| | (0.323) | (0.229) | (0.227) | (0.184) | (0.329) | (0.306) |
| Week13*ModeBrowserApple | -0.883** | -0.549* | -0.220 | -0.503** | -0.782* | -0.268 |
| | (0.332) | (0.236) | (0.233) | (0.189) | (0.339) | (0.315) |
| Week14*ModeBrowserApple | -0.572 | -0.353 | -0.283 | -0.192 | -0.491 | -0.364 |
| | (0.315) | (0.224) | (0.221) | (0.179) | (0.321) | (0.299) |
| Week15*ModeBrowserApple | -0.402 | -0.260 | 0.067 | -0.399* | -0.360 | -0.157 |
| | (0.323) | (0.229) | (0.227) | (0.184) | (0.329) | (0.306) |
| Week16*ModeBrowserApple | -0.420 | -0.039 | -0.090 | -0.261 | -0.485 | -0.134 |
| | (0.323) | (0.229) | (0.227) | (0.184) | (0.329) | (0.306) |
| Week8*ModeBrowserGoogle | -0.169 | -0.088 | 0.078 | -0.214 | -0.103 | 0.149 |
| | (0.344) | (0.244) | (0.242) | (0.196) | (0.351) | (0.326) |
| Week9*ModeBrowserGoogle | 0.079 | -0.094 | 0.083 | -0.082 | 0.035 | 0.069 |
| | (0.344) | (0.244) | (0.242) | (0.196) | (0.351) | (0.326) |
| Week10*ModeBrowserGoogle | -0.259 | -0.223 | -0.116 | -0.163 | 0.010 | -0.222 |
| | (0.303) | (0.215) | (0.213) | (0.173) | (0.310) | (0.288) |
| Week11*ModeBrowserGoogle | -0.313 | -0.263 | -0.098 | -0.168 | -0.260 | -0.216 |
| | (0.309) | (0.219) | (0.217) | (0.176) | (0.315) | (0.293) |
| Week12*ModeBrowserGoogle | -0.300 | -0.342 | -0.298 | -0.071 | -0.159 | -0.188 |
| | (0.323) | (0.230) | (0.227) | (0.184) | (0.330) | (0.307) |
| Week13*ModeBrowserGoogle | -0.696* | -0.409 | -0.171 | -0.398* | -0.612 | -0.182 |
| | (0.333) | (0.236) | (0.234) | (0.189) | (0.339) | (0.316) |
| Week14*ModeBrowserGoogle | -0.408 | -0.240 | -0.270 | -0.083 | -0.337 | -0.287 |
| | (0.316) | (0.224) | (0.222) | (0.180) | (0.322) | (0.299) |
| Week15*ModeBrowserGoogle | -0.269 | -0.173 | 0.089 | -0.315 | -0.240 | -0.096 |
| | (0.323) | (0.230) | (0.227) | (0.184) | (0.330) | (0.307) |
| Week16*ModeBrowserGoogle | -0.240 | 0.076 | -0.038 | -0.162 | -0.338 | 0.007 |

| Dependent Variables / Independent Variables | Article Views | Breadth | Visits | Article Views per Visit | Hard News | Soft News |
|---|---|---|---|---|---|---|
| | (0.323) | (0.230) | (0.227) | (0.184) | (0.330) | (0.307) |
| Week8*ModeBrowserMicrosoft | -0.148 | -0.079 | 0.064 | -0.194 | -0.080 | 0.112 |
| | (0.343) | (0.244) | (0.241) | (0.195) | (0.350) | (0.325) |
| Week9*ModeBrowserMicrosoft | -0.108 | -0.175 | 0.042 | -0.191 | -0.071 | 0.053 |
| | (0.343) | (0.244) | (0.241) | (0.195) | (0.350) | (0.325) |
| Week10*ModeBrowserMicrosoft | -0.460 | -0.335 | -0.175 | -0.284 | -0.110 | -0.240 |
| | (0.303) | (0.215) | (0.213) | (0.172) | (0.309) | (0.287) |
| Week11*ModeBrowserMicrosoft | -0.462 | -0.332 | -0.137 | -0.260 | -0.319 | -0.237 |
| | (0.308) | (0.219) | (0.217) | (0.175) | (0.315) | (0.292) |
| Week12*ModeBrowserMicrosoft | -0.450 | -0.418 | -0.316 | -0.169 | -0.244 | -0.218 |
| | (0.323) | (0.229) | (0.227) | (0.184) | (0.329) | (0.306) |
| Week13*ModeBrowserMicrosoft | -0.899** | -0.491* | -0.245 | -0.500** | -0.696* | -0.254 |
| | (0.332) | (0.236) | (0.233) | (0.189) | (0.339) | (0.315) |
| Week14*ModeBrowserMicrosoft | -0.610 | -0.309 | -0.302 | -0.203 | -0.441 | -0.327 |
| | (0.315) | (0.224) | (0.221) | (0.179) | (0.321) | (0.299) |
| Week15*ModeBrowserMicrosoft | -0.435 | -0.237 | 0.056 | -0.417* | -0.320 | -0.122 |
| | (0.323) | (0.229) | (0.227) | (0.184) | (0.329) | (0.306) |
| Week16*ModeBrowserMicrosoft | -0.439 | -0.011 | -0.101 | -0.276 | -0.456 | -0.060 |
| | (0.323) | (0.229) | (0.227) | (0.184) | (0.329) | (0.306) |
| Week8*ModeBrowserMozilla | -0.183 | -0.110 | 0.033 | -0.190 | -0.085 | 0.101 |
| | (0.344) | (0.244) | (0.242) | (0.196) | (0.351) | (0.326) |
| Week9*BrowsertypeMozilla | -0.076 | -0.192 | -0.023 | -0.108 | -0.065 | 0.003 |
| | (0.344) | (0.244) | (0.242) | (0.196) | (0.351) | (0.326) |
| Week10:BrowsertypeMozilla | -0.437 | -0.347 | -0.224 | -0.217 | -0.131 | -0.322 |
| | (0.304) | (0.216) | (0.213) | (0.173) | (0.310) | (0.288) |
| Week11*ModeBrowserMozilla | -0.442 | -0.349 | -0.192 | -0.194 | -0.351 | -0.299 |
| | (0.309) | (0.220) | (0.217) | (0.176) | (0.315) | (0.293) |
| Week12*ModeBrowserMozilla | -0.429 | -0.413 | -0.390 | -0.089 | -0.216 | -0.266 |
| | (0.324) | (0.230) | (0.227) | (0.184) | (0.330) | (0.307) |
| Week13*ModeBrowserMozilla | -0.983** | -0.576* | -0.370 | -0.477* | -0.803* | -0.375 |
| | (0.333) | (0.236) | (0.234) | (0.189) | (0.340) | (0.316) |
| Week14*ModeBrowserMozilla | -0.602 | -0.329 | -0.428 | -0.114 | -0.468 | -0.420 |
| | (0.316) | (0.224) | (0.222) | (0.180) | (0.322) | (0.300) |
| Week15*ModeBrowserMozilla | -0.458 | -0.290 | -0.057 | -0.343 | -0.406 | -0.214 |
| | (0.324) | (0.230) | (0.227) | (0.184) | (0.330) | (0.307) |
| Week16*ModeBrowserMozilla | -0.456 | -0.067 | -0.218 | -0.203 | -0.505 | -0.153 |
| | (0.324) | (0.230) | (0.227) | (0.184) | (0.330) | (0.307) |
| $R^2$ | 0.554 | 0.547 | 0.624 | 0.290 | 0.492 | 0.270 |
| N | 68,393 | 68,393 | 68,393 | 68,393 | 68,393 | 68,393 |

Notes: Each column refers to a separate regression of specification (5) on the unmatched sample. Each cell refers to the respective coefficient of the independent variables in the first column on the unmatched sample.
***p < 0.001, **p < 0.01, *p < 0.05.

# Table S6. Robustness Check on Propensity Score Matching (PSM)

| | **Article Views** | **Breadth** | **Visits** | **Article Views per Visit** | **Hard News** | **Soft News** |
|---|---|---|---|---|---|---|
| $\beta_1$ | 0.255*** | 0.166*** | 0.201*** | 0.049*** | 0.240*** | 0.145*** |
| | (0.020) | (0.013) | (0.014) | (0.009) | (0.021) | (0.018) |
| $\beta_2$ | 0.129*** | 0.082*** | 0.104*** | 0.027* | 0.162*** | 0.082*** |
| | (0.027) | (0.017) | (0.020) | (0.011) | (0.028) | (0.024) |
| N | 30,766 | 30,766 | 30,766 | 30,766 | 30,766 | 30,766 |
| $R^2$ | 0.623 | 0.590 | 0.698 | 0.556 | 0.636 | 0.621 |

Notes: $\beta_1$ represents the 1-week effect and $\beta_2$ represents the 5-week effect. Each column refers to a separate regression of the following model on a matched sample of ad blocker adopters and non-adopters using propensity score matching (psm):
$\log(Y_{it} + 1) = \alpha_i + \delta_t + \beta_1 * I_{it1}(within\ 1\ week\ of\ Treatment_{it}) + \beta_2 * I_{it2}(remaining\ weeks\ since\ Treatment_{it}) + \varepsilon_{it}$.
$R^2$ computation includes the explanatory power of the fixed effects. Standard errors clustered at the user-level appear in parentheses.
$^{***}p < 0.001$, $^{**}p < 0.01$, $^{*}p < 0.05$.

## *Web Appendix D. Robustness Checks on Different Cutoff Periods and Longer Pre-Treatment Periods*

### Table S7. Robustness Check Using 1 Week as Cutoff Period

| | Article Views | Breadth | Visits | Article Views per Visit | Hard News | Soft News |
|---|---|---|---|---|---|---|
| $\beta_1$ | 0.262*** | 0.157*** | 0.224*** | 0.035*** | 0.239*** | 0.178*** |
| | (0.008) | (0.005) | (0.006) | (0.004) | (0.008) | (0.007) |
| $\beta_2$ | 0.132*** | 0.070*** | 0.130*** | 0.010* | 0.140*** | 0.142*** |
| | (0.010) | (0.007) | (0.007) | (0.004) | (0.010) | (0.008) |
| N | 203,852 | 203,852 | 203,852 | 203,852 | 203,852 | 203,852 |
| $R^2$ | 0.689 | 0.677 | 0.747 | 0.602 | 0.686 | 0.656 |

Notes: $\beta_1$ represents the 1-week effect and $\beta_2$ represents the 5-week effect. Each column refers to a separate regression of the following model on the unmatched sample: $\log(Y_{it} + 1) = \alpha_i + \delta_t + \beta_1 * I_{it1}(within\ 1\ week\ of\ Treatment_{it}) + \beta_2 * I_{it2}(remaining\ weeks\ since\ Treatment_{it}) + \varepsilon_{it}$. $R^2$ computation includes the explanatory power of the fixed effects. Standard errors clustered at the user-level appear in parentheses. $^{***}p < 0.001$, $^{**}p < 0.01$, $^{*}p < 0.05$.

### Table S8. Robustness Check Using 3 Weeks as Cutoff Period

| | Article Views | Breadth | Visits | Article Views per Visit | Hard News | Soft News |
|---|---|---|---|---|---|---|
| $\beta_1$ | 0.240*** | 0.136*** | 0.205*** | 0.031*** | 0.203*** | 0.191*** |
| | (0.011) | (0.007) | (0.008) | (0.005) | (0.012) | (0.010) |
| $\beta_2$ | 0.083*** | 0.040*** | 0.073*** | 0.011 | 0.089*** | 0.113*** |
| | (0.014) | (0.009) | (0.010) | (0.006) | (0.014) | (0.011) |
| N | 167,668 | 167,668 | 167,668 | 167,668 | 167,668 | 167,668 |
| $R^2$ | 0.678 | 0.672 | 0.734 | 0.604 | 0.677 | 0.641 |

Notes: $\beta_1$ represents the 1-week effect and $\beta_2$ represents the 5-week effect. Each column refers to a separate regression of the following model on the unmatched sample: $\log(Y_{it} + 1) = \alpha_i + \delta_t + \beta_1 * I_{it1}(within\ 1\ week\ of\ Treatment_{it}) + \beta_2 * I_{it2}(remaining\ weeks\ since\ Treatment_{it}) + \varepsilon_{it}$. $R^2$ computation includes the explanatory power of the fixed effects. Standard errors clustered at the user-level appear in parentheses. $^{***}p < 0.001$, $^{**}p < 0.01$, $^{*}p < 0.05$.

### Table S9. Robustness Check Using 4 Weeks as Cutoff Period

| | Article Views | Breadth | Visits | Article Views per Visit | Hard News | Soft News |
|---|---|---|---|---|---|---|
| $\beta_1$ | 0.268*** | 0.152*** | 0.215*** | 0.040*** | 0.255*** | 0.190*** |
| | (0.019) | (0.013) | (0.013) | (0.009) | (0.019) | (0.017) |
| $\beta_2$ | 0.093** | 0.047* | 0.050* | 0.033* | 0.122*** | 0.097*** |
| | (0.029) | (0.020) | (0.020) | (0.014) | (0.031) | (0.027) |
| N | 142,074 | 142,074 | 142,074 | 142,074 | 142,074 | 142,074 |
| $R^2$ | 0.672 | 0.670 | 0.720 | 0.609 | 0.673 | 0.629 |

Notes: $\beta_1$ represents the 1-week effect and $\beta_2$ represents the 5-week effect. Each column refers to a separate regression of the following model on the unmatched sample: $\log(Y_{it} + 1) = \alpha_i + \delta_t + \beta_1 * I_{it1}(within\ 1\ week\ of\ Treatment_{it}) + \beta_2 * I_{it2}(remaining\ weeks\ since\ Treatment_{it}) + \varepsilon_{it}$. $R^2$ computation includes the explanatory power of the fixed effects. Standard errors clustered at the user-level appear in parentheses. $^{***}p < 0.001$, $^{**}p < 0.01$, $^{*}p < 0.05$.

## Table S10. Robustness Check Using Week 1 to Week 11 as Pre-Treatment Period

| | Article Views | Breadth | Visits | Article Views per Visit | Hard News | Soft News |
|---|---|---|---|---|---|---|
| $\beta_1$ | 0.301*** | 0.234*** | 0.268*** | 0.149*** | 0.232*** | 0.167*** |
| | (0.040) | (0.029) | (0.031) | (0.019) | (0.039) | (0.033) |
| $\beta_2$ | 0.146** | 0.087* | 0.130*** | -0.013 | 0.139** | 0.056 |
| | (0.047) | (0.035) | (0.039) | (0.020) | (0.044) | (0.031) |
| N | 14,273 | 14,273 | 14,273 | 14,273 | 14,273 | 14,273 |
| $R^2$ | 0.472 | 0.469 | 0.537 | 0.330 | 0.471 | 0.461 |

Notes: $\beta_1$ represents the 1-week effect and $\beta_2$ represents the 5-week effect. Each column refers to a separate regression of the following model on a matched sample from week 1 to week 16 (full observation period): $\log(Y_{it} + 1) = \alpha_i + \delta_t + \beta_1 * I_{it1}(within\ 1\ week\ of\ Treatment_{it}) + \beta_2 * I_{it2}(remaining\ weeks\ since\ Treatment_{it}) + \varepsilon_{it}$
$R^2$ computation includes the explanatory power of the fixed effects. Standard errors clustered at the user-level appear in parentheses. $^{***}p < 0.001$, $^{**}p < 0.01$, $^{*}p < 0.05$.

*Web Appendix E. Robustness Checks on Effect Decomposition Using Other Quasi-Experimental Designs*

**Table S11. Robustness Check on Effect Decomposition Using Ad Blocker Early Adopters as Treatment Group and Ad Blocker Late Adopters as Control Group**

| | Article Views | Breadth | Visits | Article Views per Visit | Hard News | Soft News |
|---|---|---|---|---|---|---|
| $\beta_1$ | 0.322* | 0.207* | 0.072 | 0.211** | 0.272* | 0.092 |
| | (0.127) | (0.082) | (0.059) | (0.079) | (0.126) | (0.098) |
| $\beta_2$ | -0.010 | 0.019 | 0.093 | -0.037 | 0.060 | -0.037 |
| | (0.133) | (0.086) | (0.092) | (0.058) | (0.130) | (0.110) |
| N | 1,423 | 1,423 | 1,423 | 1,423 | 1,423 | 1,423 |
| $R^2$ | 0.462 | 0.423 | 0.588 | 0.478 | 0.550 | 0.557 |

Notes: $\beta_1$ represents the 1-week effect and $\beta_2$ represents the 5-week effect. Each column refers to a separate regression of the following model on the matched sample of early and late adopters. $R^2$ computation includes the explanatory power of the fixed effects:
$\log(Y_{it}+1) = \alpha_i + \delta_t + \beta_1 * I_{it1}(within\ 1\ week\ of\ Treatment_{it}) + \beta_2 * I_{it2}(remaining\ weeks\ since\ Treatment_{it}) + \varepsilon_{it}$
Standard errors clustered at the user-level appear in parentheses. $^{***}p < 0.001$, $^{**}p < 0.01$, $^{*}p < 0.05$.

**Table S12. Robustness Check on Effect Decomposition Using Ad Blocker Abandoners as Treatment Group and Continuous Ad Blocker Users as Control Group**

| | Article Views | Breadth | Visits | Article Views per Visit | Hard News | Soft News |
|---|---|---|---|---|---|---|
| $\beta_1$ | -0.204*** | -0.127*** | -0.161*** | -0.038*** | -0.195*** | -0.063*** |
| | (0.016) | (0.010) | (0.011) | (0.007) | (0.016) | (0.013) |
| $\beta_2$ | -0.115*** | -0.076*** | -0.109*** | -0.017 | -0.107*** | 0.019 |
| | (0.023) | (0.014) | (0.016) | (0.009) | (0.023) | (0.020) |
| N | 48,833 | 48,833 | 48,833 | 48,833 | 48,833 | 48,833 |
| $R^2$ | 0.748 | 0.702 | 0.790 | 0.654 | 0.743 | 0.725 |

Notes: $\beta_1$ represents the 1-week effect and $\beta_2$ represents the 5-week effect. Each column refers to a separate regression of the following model on the unmatched sample of abandoners and continuous ad blocker users.: $\log(Y_{it}+1) = \alpha_i + \delta_t + \beta_1 * I_{it1}(within\ 1\ week\ of\ Treatment_{it}) + \beta_2 * I_{it2}(remaining\ weeks\ since\ Treatment_{it}) + \varepsilon_{it}$. $R^2$ computation includes the explanatory power of the fixed effects. Standard errors clustered at the user-level appear in parentheses. $^{***}p < 0.001$, $^{**}p < 0.01$, $^{*}p < 0.05$.

*Web Appendix F. Causal Assumptions, Challenges, and Solutions*

To identify the causal effect of ad blocker adoption on news consumption, we follow the Rubin Causal Model, the workhorse model in causal inference in the economics and marketing literature (Imbens and Wooldridge 2009; Rosenbaum and Rubin 1983). Imbens and Wooldridge (2009) summarize the causal assumptions under the Rubin Causal Model as follows:

Assumption 1. Unconfoundedness

$$w_i \perp (Y_i(0), Y_i(1)) | X_i.$$

The unconfoundedness assumption assumes that no observed or unobserved variables ($X_i$) correlate with the potential outcome ($Y_i(0)\ or\ Y_i(1)$) and the treatment ($w_i$).

Assumption 2. Overlap

$$0 < pr(w_i = 1 | X_i = x) < 1, for\ all\ x.$$

The overlap assumption assumes that the support of the conditional distribution of $X_i$ given $w_i = 0$ overlaps completely with that of the conditional distribution of $X_i$ given $w_i =$ 1.

We summarize these causal assumptions as well as the resulting identification challenges and our solutions in Table S13.

**Table S13. Causal Assumptions, Challenges, and Solutions**

| Causal Assumption | Challenge | Solution | Specification |
|---|---|---|---|
| Unconfoundedness | Users differ in observed and time-invariant ways | Coarsened exact matching | Main Specification (Table 4) |
| | Users differ in unobserved and time-invariant ways | Individual-fixed effect | |
| | Seasonality | Week-fixed effect | |
| | Users differ in unobserved reasons related to ad blocker adoption | Compare results for subsamples of ad blocker users (i.e., early vs. late adopters and abandoners vs. always adopters) | Robustness Check (Table 4) |
| | Unexplained part of adoption decision (error term) related to news consumption | Heckman selection model | Robustness Check (Table S14 & S15) |
| | Users differ in time-varying ways | Time-varying controls | Robustness Check (Table S2 & S5) |
| Overlap | Unbalanced empirical distribution of covariates between treatment and control group | Coarsened exact matching | Table 3 |

*Web Appendix G. Heckman Selection Model*

To formally test whether a user's ad blocker adoption poses a selection bias (e.g., a user anticipates to read more news and thus adopts an ad blocker), we use the Heckman selection model (Heckman 1979). In the marketing literature, the Heckman selection model has been used in the past to test and correct for the selection bias of an individual's (a household's) technology adoption decision (Bronnenberg et al. 2010; Narang and Shankar 2019).

In the first stage, we model a user's ad blocker adoption decision. Survey studies (Mathur et al. 2018; Newman et al. 2016; Pritchard 2021; Redondo and Aznar 2018; Singh and Potdar 2009; Sołtysik-Piorunkiewicz et al. 2019; Vratonjic et al. 2013) outline that users adopt an ad blocker mainly for three reasons: the annoyance of ads, page loading speed, and privacy concerns. Therefore, we look for user characteristics that can serve as proxies for these three reasons to adopt an ad blocker.

First, to proxy for ad annoyance, we use the user's number of page views from mobile (rather than desktop) devices. We assume that ads are likely to be more annoying on mobile devices because mobile devices tend to have smaller screen sizes than desktops. Second, we capture page loading speed by a user's JavaScript version and assume that users with older JavaScript versions load pages slower than users with newer JavaScript versions. Third, for the role of privacy concerns, we use a binary indicator that is one if a user has ever rejected a cookie (and zero otherwise). The underlying assumption is that users who have ever rejected a cookie are more privacy-concerned than users who never rejected a cookie.

In addition, we add a user's most frequently-used browser in the pre-treatment period as an explanatory variable into the model for the ad blocker adoption decision. The reason is that an external event happened during our observation period that triggered a large amount of consumer awareness and a significant increase in the adoption of ad blockers. That event is the release of iOS9 on the Apple iPhone on September 16, 2015, which is in week 15 of our observation period. With iOS9, Apple offered, for the first time, a mobile operating system that allowed for a content blocking feature. That feature made mobile ad blocker software for iPhone possible and also created massive consumer awareness for ad blockers. Google

Search Trends support this massive increase in consumer behavior because the search volume for the term “ad blocker” peaks to its largest volume ever (i.e., 100) during September 2015, as shown in Figure S1 and Figure S2.

When we check the related queries for the keyword “ad blocker” during September 2015, we find that the search query with the largest increase is “best iOS 9 ad blocker” and the most popular search query is “chrome ad blocker”, as shown in Figure S3 and Figure S4. The most popular search query reflects that users also learn about ad blockers for other browsers (here: Chrome) during the iOS 9 event. All those observations are in line with our expectation that the release of iOS9 yields a high ad blocker adoption rate during our observation period. Thus, we include a variable that captures a user’s most frequently used browser in the pre-treatment period (for short: mode browser) as the exclusive variables for the ad blocker adoption decision. Our idea is that a user who most frequently used an Apple browser in the pre-treatment period is more likely to adopt an ad blocker because of the external event in which Apple made an ad blocking feature available.

**Figure S1. Google Search for the Term “Ad Blocker” (Worldwide) from 2004 to 2020**

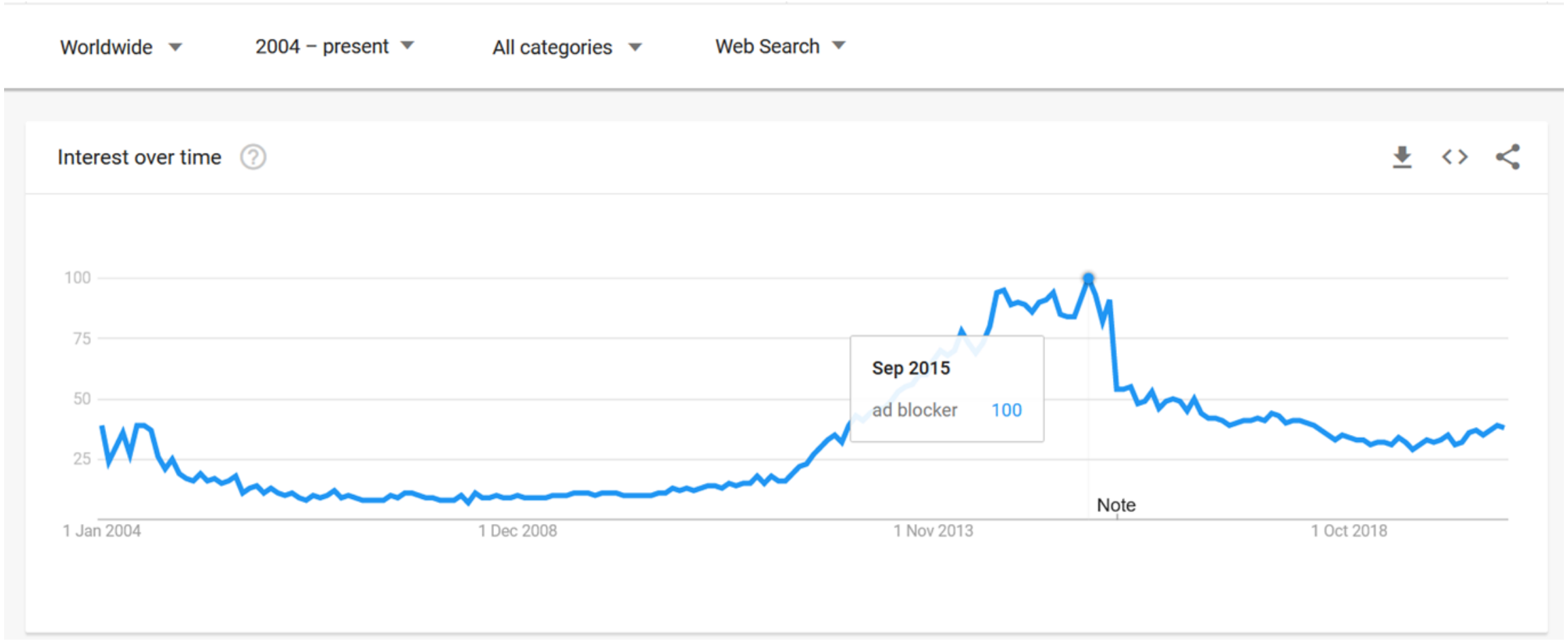

**Figure S2. Google Search for the Term "Ad Blocker" (Worldwide) in September, 2015**

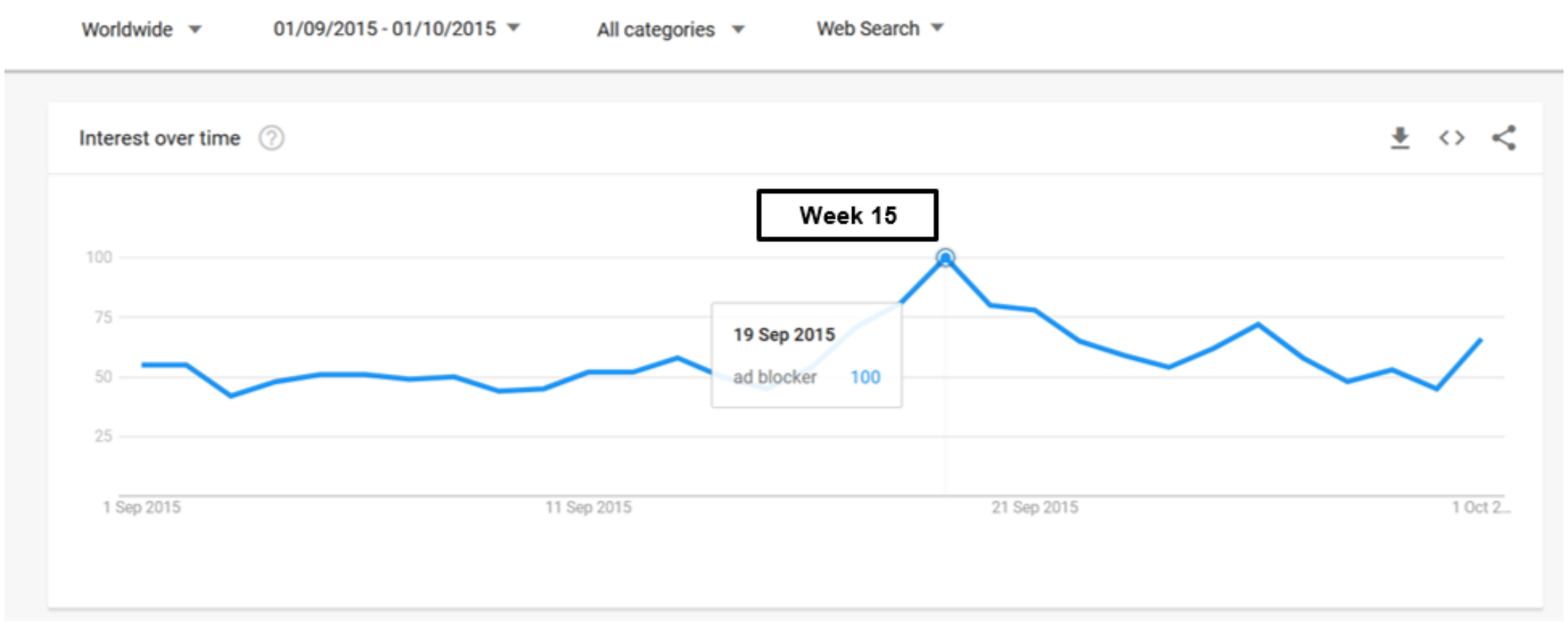

**Figure S3. Related Topics and Related Queries with the Biggest Increase for "Ad Blocker" on Google Search Trends (Worldwide) in September, 2015**

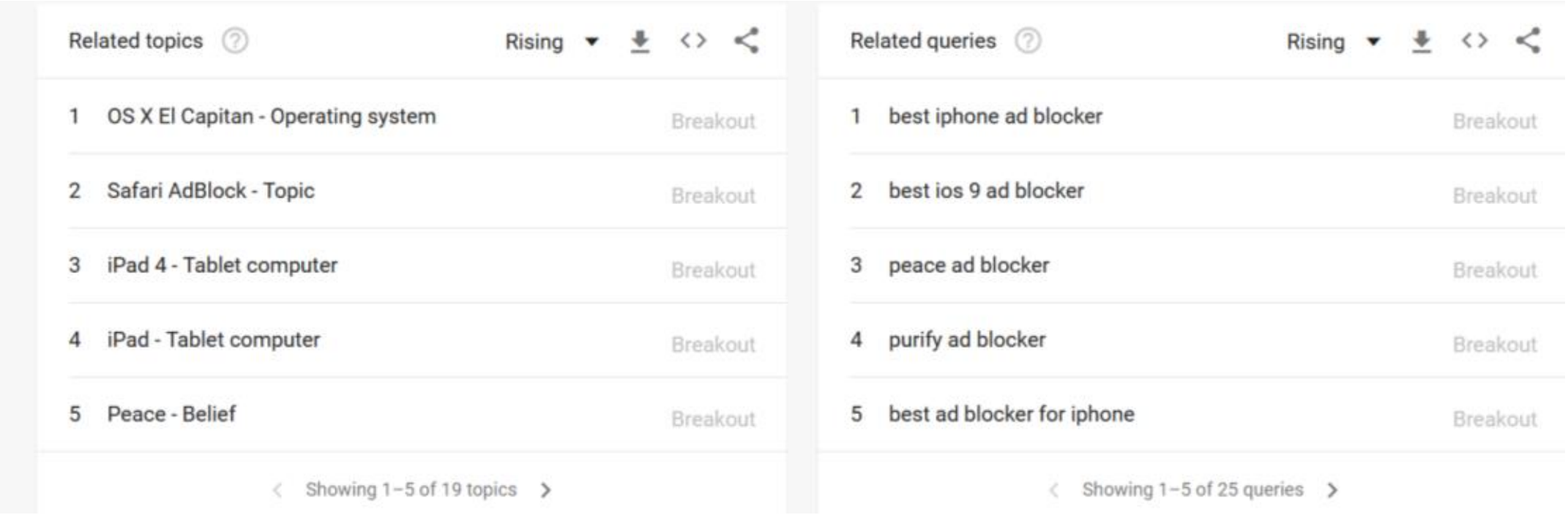

**Figure S4. Most Popular Related Topics and Related Queries for "Ad Blocker" on Google Search Trends (Worldwide) in September, 2015**

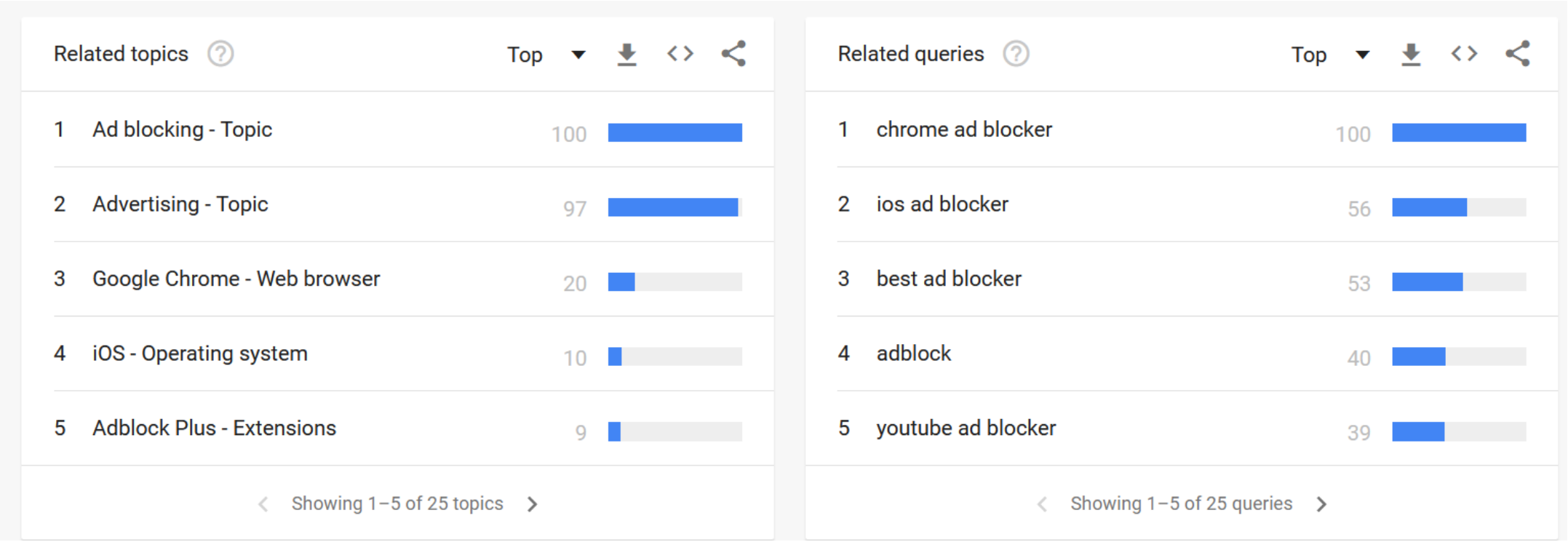

Thus, in the first stage, we estimate the following probit model:

(6) $$Adoption_i = \alpha + \beta_1 * CookieDeleted_i + \beta_2 * MobileViews_i + \beta_3 * JavaScript_i + \beta_4 * ModeBrowser_i + \varepsilon_i,$$

where $Adoption_i$ is the ad blocker adoption decision of a user $i$; $\alpha$ is the intercept; $CookieDeleted_i$ is coded as 1 if user $i$ has not always accepted a cookie in the pre-treatment period; $MobileViews_i$ is a user's average weekly number of page impressions generated on a mobile device in the pre-treatment period; $JavaScript_i$ is the most frequently used JavaScript version of a user $i$ in the pre-treatment period, with a version below 1.5 as baseline; $ModeBrowser_i$ is a dummy-coded categorical variable that describes whether a particular browser (which could be either Apple Safari, Google Chrome, Microsoft Internet Explorer, Mozilla Firefox, or a group of all the other browsers) is the most frequently used browser by user $i$ in the pre-treatment period, with Microsoft Internet Explorer as a baseline because it is the default browser for Windows and most likely used by the least tech-savvy users; $\varepsilon_i$ is the error term.

**Table S14. First Stage of Heckman Selection Model**

| | Adoption |
|---|---|
| (Intercept) | 3.800 |
| | (48.178) |
| CookieDeleted | 0.379 |
| | (0.403) |
| MobileVisits | $0.012^{***}$ |
| | (0.001) |
| JavaScript1.5 | -7.857 |
| | (68.134) |
| JavaScript1.6 | -4.543 |
| | (48.178) |
| JavaScript1.8 | -4.055 |
| | (48.178) |
| ModeBrowserApple | $0.831^{***}$ |
| (Baseline: Microsoft) | (0.022) |
| ModeBrowserGoogle | $0.257^{***}$ |
| | (0.028) |
| ModeBrowserMozilla | $0.498^{***}$ |
| | (0.043) |
| ModeBrowserOthers | $2.035^{***}$ |
| | (0.480) |
| N | 21,068 |

Notes: The model is estimated on a sample not matched with the first stage variables but matched with all other variables in Table 3. $^{***}p < 0.001$, $^{**}p < 0.01$, $^{*}p < 0.05$. Standard errors appear in parentheses.

Table S14 reports the result of the first stage of the Heckman selection model. According to our expectation, users who are more likely to browse on a mobile device and use non-Microsoft browsers most frequently (i.e., either Apple Safari, Google Chrome, Mozilla Firefox or any other browsers) are more likely to adopt an ad blocker.

In the second stage, we estimate the same model (1) as in the main paper. Additionally, we include the inverse mills ratio (from the first stage of the Heckman selection model, see Table S14), capturing that the first-stage error could affect the second-stage model. Notably, any variables omitted by the first stage would be captured in the error term, including whether a user anticipates reading more news or spending more time online. A t-test on the inverse mills ratio is thus a direct test on the selection bias of ad blocker adoption on news consumption while making only one joint normality assumption of the first and second stage error terms.

Table S15 reports the second-stage results of Heckman selection model.

**Table S15. Second-Stage Results of Heckman Selection Model**

| | Article Views | Breadth |
|---|---|---|
| (Intercept) | 1.655*** | 1.303*** |
| | (0.036) | (0.025) |
| $\beta_1$ | 0.259** | 0.147 |
| | (0.032) | (0.023) |
| $\beta_2$ | 0.171*** | 0.126*** |
| | (0.027) | (0.019) |
| invMillsRatio | -0.073 | -0.019 |
| | (0.039) | (0.028) |
| Sigma | 0.978 | 0.684 |
| Rho | -0.075 | -0.027 |
| N | 21,068 | 21,068 |

Note: The model is estimated on a sample not matched with the first stage variables but matched with all other variables in Table 3.
***p < 0.001, **p < 0.01, *p < 0.05. Standard errors appear in parentheses.

As shown in Table S15, the 1-week and 5-week effects ($\beta_1$ and $\beta_2$) are still significant and qualitatively similar to our estimates from the main specification (shown in Table 4). More importantly, the parameter of the inverse Mills Ratio is -0.073 with a t-value of 0.039 for article views and -0.019 with a t-value of 0.028, suggesting that we cannot reject the null hypothesis (i.e., the absence of a sample selection bias) at our conventional significance level ($p < 0.05$). In other words, we do not find evidence for selection bias in our case (e.g., it is not a user's anticipated news consumption influences his/her decision to adopt an ad blocker, especially considering that ad blockers perform the same function on all websites, whether or not it's a news website or not).

*Web Appendix H. Robustness Checks on Logarithm Transformed Dependent Variable*

**Table S16. Robustness Check on Using Original Value (Instead of Log) as Dependent Variable**

| | Ad Blocker Adoption | | Ad Blocker Early Adoption | | Ad Blocker Abandonment | |
|---|---|---|---|---|---|---|
| | **Article Views** | **Breadth** | **Article Views** | **Breadth** | **Article Views** | **Breadth** |
| $\beta_1$ | 2.309*** | 0.873*** | 3.126** | 1.039** | -1.350 | -0.530 |
| | (0.356) | (0.111) | (1.192) | (0.340) | (1.215) | (0.377) |
| $\beta_2$ | 1.419*** | 0.425** | 0.441 | 0.214 | 0.281 | 0.168 |
| | (0.412) | (0.143) | (1.473) | (0.383) | (1.998) | (0.490) |
| N | 9,370 | 9,370 | 1,423 | 1,423 | 1,009 | 1,009 |
| $R^2$ | 0.503 | 0.497 | 0.477 | 0.442 | 0.554 | 0.570 |

Notes: $\beta_1$ represents the 1-week effect and $\beta_2$ represents the 5-week effect. Each column refers to a separate regression of the following model on the matched sample: $Y_{it} = \alpha_i + \delta_t + \beta_1 * I_{it1}(within\ 1\ week\ of\ Treatment_{it}) + \beta_2 * I_{it2}(remaining\ weeks\ since\ Treatment_{it}) + \varepsilon_{it}$. $R^2$ computation includes the explanatory power of the fixed effects. Standard errors clustered at the user-level appear in parentheses. ***p < 0.001, **p < 0.01, *p < 0.05.

**Table S17. Robustness Check on Using Log (Y + 0.1) as Dependent Variable**

| | Ad Blocker Adoption | | Ad Blocker Early Adoption | | Ad Blocker Abandonment | |
|---|---|---|---|---|---|---|
| | **Article Views** | **Breadth** | **Article Views** | **Breadth** | **Article Views** | **Breadth** |
| $\beta_1$ | 0.434*** | 0.315*** | 0.450* | 0.312* | -0.420* | -0.250 |
| | (0.063) | (0.051) | (0.197) | (0.149) | (0.187) | (0.151) |
| $\beta_2$ | 0.232** | 0.161* | -0.041 | -0.002 | 0.134 | 0.127 |
| | (0.083) | (0.069) | (0.198) | (0.152) | (0.218) | (0.172) |
| N | 9,370 | 9,370 | 1,423 | 1,423 | 1,009 | 1,009 |
| $R^2$ | 0.465 | 0.455 | 0.425 | 0.384 | 0.524 | 0.507 |

Notes: $\beta_1$ represents the 1-week effect and $\beta_2$ represents the 5-week effect. Each column refers to a separate regression of the following model on the matched sample: $\log(Y_{it} + 0.1) = \alpha_i + \delta_t + \beta_1 * I_{it1}(within\ 1\ week\ of\ Treatment_{it}) + \beta_2 * I_{it2}(remaining\ weeks\ since\ Treatment_{it}) + \varepsilon_{it}$. $R^2$ computation includes the explanatory power of the fixed effects. Standard errors clustered at the user-level appear in parentheses. ***p < 0.001, **p < 0.01, *p < 0.05.

## *Web Appendix I. Treatment Effect on Article Views in Each News Category and on Page Views in Each Non-News Category*

### Table S18. Treatment Effect on Article Views in Each News Category

| | International Political | Regional Political | Local Political | Economy | Finance | Opinion | Sport | Art & Culture |
|---|---|---|---|---|---|---|---|---|
| $\beta_1$ | 0.184*** | 0.108*** | 0.068** | 0.145*** | 0.086*** | 0.085*** | 0.120*** | 0.061*** |
| | (0.032) | (0.028) | (0.021) | (0.027) | (0.024) | (0.022) | (0.028) | (0.018) |
| $\beta_2$ | 0.092* | 0.055 | 0.046 | 0.088** | 0.065* | 0.056* | 0.034 | 0.030 |
| | (0.040) | (0.034) | (0.027) | (0.031) | (0.028) | (0.027) | (0.030) | (0.022) |
| N | 9,370 | 9,370 | 9,370 | 9,370 | 9,370 | 9,370 | 9,370 | 9,370 |
| $R^2$ | 0.485 | 0.454 | 0.425 | 0.437 | 0.575 | 0.346 | 0.594 | 0.348 |

| | Lifestyle | Brief | News Ticker | Outlook | Transportat ion | Science | Sunday News | Photo stream |
|---|---|---|---|---|---|---|---|---|
| $\beta_1$ | 0.015 | 0.009 | 0.016 | 0.078*** | 0.010 | 0.025 | 0.030* | 0.021 |
| | (0.011) | (0.006) | (0.016) | (0.021) | (0.007) | (0.013) | (0.015) | (0.012) |
| $\beta_2$ | 0.013 | 0.013 | -0.007 | 0.035 | 0.019 | -0.004 | 0.014 | 0.005 |
| | (0.014) | (0.009) | (0.016) | (0.026) | (0.014) | (0.013) | (0.021) | (0.014) |
| N | 9,370 | 9,370 | 9,370 | 9,370 | 9,370 | 9,370 | 9,370 | 9,370 |
| $R^2$ | 0.314 | 0.265 | 0.471 | 0.368 | 0.348 | 0.367 | 0.266 | 0.304 |

| | Video | Digital | Special | Data |
|---|---|---|---|---|
| $\beta_1$ | 0.013* | 0.041** | 0.004 | 0.000 |
| | (0.006) | (0.015) | (0.004) | (0.002) |
| $\beta_2$ | 0.006 | 0.015 | 0.003 | 0.001 |
| | (0.008) | (0.018) | (0.003) | (0.003) |
| N | 9,370 | 9,370 | 9,370 | 9,370 |
| $R^2$ | 0.136 | 0.455 | 0.160 | 0.193 |

Notes: $\beta_1$ represents the 1-week effect and $\beta_2$ represents the 5-week effect. Each column refers to a separate regression of the following model on the matched sample of ad blocker adopters and non-adopters : $Y_{it} = \alpha_i + \delta_t + \beta_1 * I_{it1}(within\ 1\ week\ of\ Treatment_{it}) + \beta_2 * I_{it2}(remaining\ weeks\ since\ Treatment_{it}) + \varepsilon_{it}$. $R^2$ computation includes the explanatory power of the fixed effects. Standard errors clustered at the user-level appear in parentheses. ***p < 0.001, **p < 0.01, *p < 0.05.

### Table S19. Treatment Effect on Page Views in Each Non-News Category

| | Homepage | Weather | Play Page | Account | Others | Search | Archive |
|---|---|---|---|---|---|---|---|
| $\beta_1$ | 0.258*** | 0.016 | 0.002 | 0.006 | 0.015 | 0.003 | 0.008* |
| | (0.035) | (0.015) | (0.005) | (0.022) | (0.011) | (0.014) | (0.004) |
| $\beta_2$ | 0.137** | 0.012 | 0.001 | 0.000 | 0.012 | -0.015 | 0.006 |
| | (0.044) | (0.018) | (0.008) | (0.030) | (0.013) | (0.018) | (0.003) |
| N | 9,370 | 9,370 | 9,370 | 9,370 | 9,370 | 9,370 | 9,370 |
| $R^2$ | 0.665 | 0.709 | 0.759 | 0.332 | 0.207 | 0.283 | 0.191 |

Notes: $\beta_1$ represents the 1-week effect and $\beta_2$ represents the 5-week effect. Each column refers to a separate regression of the following model on the matched sample of ad blocker adopters and non-adopters: $Y_{it} = \alpha_i + \delta_t + \beta_1 * I_{it1}(within\ 1\ week\ of\ Treatment_{it}) + \beta_2 * I_{it2}(remaining\ weeks\ since\ Treatment_{it}) + \varepsilon_{it}$. $R^2$ computation includes the explanatory power of the fixed effects. Standard errors clustered at the user-level appear in parentheses. ***p < 0.001, **p < 0.01, *p < 0.05.

*Web Appendix J. Robustness Check on Zero Visit Weeks*

**Table S20. Robustness on Article Views and Breadth with Zero Visit Weeks**

| | Article Views | Breadth |
|---|---|---|
| $\beta_1$ | 0.510*** | 0.461*** |
| | (0.037) | (0.029) |
| $\beta_2$ | 0.254*** | 0.248*** |
| | (0.048) | (0.038) |
| N | 13,220 | 13,220 |
| $R^2$ | 0.531 | 0.615 |

Notes: $\beta_1$ represents the 1-week effect and $\beta_2$ represents the 5-week effect. Each column refers to a separate regression of the following model on the matched sample of ad blocker adopters and non-adopters with user zero visit weeks included: $\log(Y_{it}+1) = \alpha_i + \delta_t + \beta_1 * I_{it1}(within\ 1\ week\ of\ Treatment_{it}) + \beta_2 * I_{it2}(remaining\ weeks\ since\ Treatment_{it}) + \varepsilon_{it}$. ***p < 0.001, **p < 0.01, *p < 0.05

**Table S21. Robustness on Article Views and Breadth with Tobit Model**

| | Article Views | Breadth |
|---|---|---|
| $\beta_1$ | 0.362*** | 0.221*** |
| | (0.037) | (0.024) |
| $\beta_2$ | 0.205*** | 0.118*** |
| | (0.044) | (0.029) |
| Log Likelihood | -7689.2 | -5240.8 |
| N | 9,370 | 9,370 |

Notes: $\beta_1$ represents the 1-week effect and $\beta_2$ represents the 5-week effect. Each column refers to a separate regression of the following model on the matched sample without user zero visit week using a truncated regression model: $\log(Y_{it}+1) = \alpha_i + \delta_t + \beta_1 * I_{it1}(within\ 1\ week\ of\ Treatment_{it}) + \beta_2 * I_{it2}(remaining\ weeks\ since\ Treatment_{it}) + \varepsilon_{it}$. ***p < 0.001, **p < 0.01, *p < 0.05

## *Web Appendix K. Examination of User Demographics on Ad Blocker Adoption*

**Table S22. Logistic Regression of Ad Blocker Adoption on User Demographics**

| | **Ad Blocker Adoption** |
|---|---|
| (Intercept) | -1.669*** |
| | (0.281) |
| Gender (Male) | 0.282*** |
| | (0.058) |
| Income Index 2 | -0.456*** |
| | (0.127) |
| Income Index 3 | -0.236* |
| | (0.115) |
| Income Index 4 | -0.347** |
| | (0.130) |
| Income Index 5 | -0.352** |
| | (0.116) |
| Income Index 6 | -0.259* |
| | (0.114) |
| Age 25 – 19 | 0.251 |
| | (0.324) |
| Age 30 – 34 | 0.240 |
| | (0.303) |
| Age 35 – 39 | 0.405 |
| | (0.281) |
| Age 40 – 44 | 0.316 |
| | (0.273) |
| Age 45 – 49 | 0.354 |
| | (0.270) |
| Age 50 – 54 | 0.239 |
| | (0.270) |
| Age 55 – 59 | 0.107 |
| | (0.271) |
| Age 60 – 64 | 0.316 |
| | (0.270) |
| Age 65 – 69 | -0.004 |
| | (0.273) |
| Age 70 – 74 | -0.016 |
| | (0.273) |
| Age 75 – 79 | -0.014 |
| | (0.279) |
| Age 80 – 85 | 0.228 |
| | (0.282) |
| AIC | 13161.475 |
| BIC | 13305.086 |
| Log Likelihood | -6561.738 |
| Deviance | 13123.475 |
| N | 14,164 |

Notes: This table reports the coefficient of the following logit model on the unmatched sample: $Adoption_i = \alpha + \beta_1 * Gender_i + \beta_2 * Income_i + \beta_3 * Age_i + \varepsilon_i$; exp (β) is the odds ratio between ad blocker adopters and non-adopters. The reference group for gender is female, for income is income index 1 (the lowest income category), for age is 20-24. The coefficient of Gender (male) indicates that the odds of being an ad blocker adopter in the male group is exp (0.282) = 1.33 times that of being an ad blocker in the female group. The coefficient of Income Index 2 indicates that the odds of being an ad blocker adopter in the Income Index 2 group is exp (0.282) = 0.63 times that of being an ad blocker in the Income Index 1 group. $^{***}p < 0.001$, $^{**}p < 0.01$, $^{*}p < 0.05$

Heidelberg: Springer, 49-73.